\documentclass[twocolumn]{aastex631}
\usepackage{graphicx}
\usepackage{xcolor}
\usepackage [english]{babel}
\usepackage [autostyle, english = american]{csquotes}
\MakeOuterQuote{"}

\begin{document}

\title{Radio AGN Activity in Low Redshift Galaxies is Not Directly Related to Star Formation Rates}

\author{Arjun Suresh}
\author{Michael R. Blanton}
\affil{New York University}


\begin{abstract}
 We examine the demographics of radio-emitting
 active galactic nuclei (AGN)
 in the local universe as a function
 of host galaxy properties, most notably both stellar mass and star formation rate. 
 Radio AGN activity is theoretically implicated in helping reduce star formation
 rates of galaxies, and therefore it is natural to investigate the relationship
 between these two galaxy properties.  
 We use a sample of $\sim 10^4$ 
 galaxies from the Mapping Nearby Galaxies at APO (MaNGA)
 survey, part of the Sloan Digital Sky Survey IV (SDSS-IV),
 along with the Faint Images of the Radio Sky at Twenty 
 centimeters (FIRST) radio survey and the National Radio Astronomy Observatory (NRAO) Very Large Array (VLA) 
Sky Survey (NVSS). There are $1,126$ galaxies in MaNGA with radio detections.
 Using star formation
 rate and stellar mass estimates based on Pipe3D, 
 inferred from the high 
 signal-to-noise ratio measurements
 from MaNGA, we show that star formation rates
 are strongly correlated with 20~cm radio
 emission, as expected. We 
 identify as radio AGN those radio emitters that are 
 much stronger  than expected from the star formation rate.
 Using this sample of AGN, the well-measured
 stellar velocity dispersions from MaNGA, and the 
 black hole $M$-$\sigma$ relationship, we examine
 the Eddington ratio distribution and its dependence
 on stellar mass and star formation rate. 
 We find that the 
 Eddington ratio distribution depends strongly on stellar mass,
 with more massive galaxies having larger Eddington ratios.
 Interpreting our model fit to the data leads to a completeness-corrected
 estimate of $F_{\rm AGN}(\lambda>0.01)$, the fraction of galaxies
 with radio AGN with an Eddington ratio $\lambda > 0.01$.
 At $\log\left(M_\star/M_{\odot}\right) \sim 11$, we estimate 
 $F_{\rm AGN} = 0.03$. 
 As found in previous studies, the AGN fraction
 increases rapidly with $M_\star$, and  at  $\log\left(M_\star/M_{\odot}\right) \sim 12$ 
 we estimate  $F_{\rm AGN}\sim 0.36$.
 We do not find any dependence on star formation rate, specific star formation rate,
 or velocity dispersion when controlling for stellar mass.
 We conclude that galaxy star formation rates appear 
 to be unrelated to the presence or absence of a radio AGN, which may be 
 useful in constraining theoretical models of AGN feedback. 
\end{abstract}

\section{Introduction}

Active Galactic Nuclei (AGN) are supermassive black holes 
at the centers of galaxies, shining brightly due to rapid 
accretion of matter, and detectable across
the electromagnetic spectrum. In this paper, we will study
radio emitting AGN, which have synchrotron emission
that can dominate the mm and cm wavelength spectrum. 

\citet{matthews64a} first noted that radio AGN tend to
be in giant elliptical galaxies (specifically in 
``D'' galaxies in Morgan's classification). \cite{heckman14a} 
describe and present the modern evidence confirming this
tendency. That review largely builds off of the work of 
\cite{best05a} and \cite{best12a} with the
Sloan Digital Sky Survey I and II (SDSS-I, SDSS-II; \citealt{york00a}),
the National Radio Astronomy Observatory (NRAO) Very Large Array (VLA) 
Sky Survey (NVSS; \citealt{condon98a}),
and the Faint Images of the Radio Sky
at Twenty centimeters survey (FIRST; \citealt{becker95a}).

Those investigations quantified the radio luminosity function
as a function of galaxy stellar mass, and showed that the radio
luminosity function is a strong function of mass, such that 
the ratio of radio luminosity to mass grows with mass.
\citet{hickox09a} reported similar effects for radio 
AGN at $z\sim0.25$---$0.8$ from the AGN and Galaxy Evolution Survey 
(AGES; \citealt{kochanek12a}).
This tendency holds for the most luminous radio galaxies (e.g. 
those with $\nu L_{\nu, {\rm 1.4 GHz}}$ above about $10^{41}$ erg s$^{-1}$)
which are commonly visible also in optical broad lines,
but also in the more moderately luminous radio galaxies.

Partly based on these observations, a theoretical picture
has emerged that explains the lack of significant star 
formation in the most  massive galaxies in the universe 
as a result of the ``radio-mode'' feedback from the moderately
luminous radio jets. 
Early work implementing radio-mode feedback 
included \citet{croton06a}, \citet{bower06a}, and
\citet{sijacki07a}, but for 
more comprehensive reviews see
\citet{fabian12a}, \citet{somerville15a},
and \citet{vogelsberger20a}. 

In this paper, we revisit the demographics of radio AGN
using the integral field survey observations from 
Mapping Nearby Galaxies at APO (MaNGA; \citealt{bundy15a}),
a component of the Sloan Digital Sky Survey IV (SDSS-IV;
\citealt{blanton17a}). MaNGA has a sample of $\sim 10^4$ galaxies,
much smaller than the SDSS-I and -II samples, but with 
much more reliable and higher signal-to-noise ratio 
determinations of galaxy properties, including stellar masses,
star formation rates, and velocity dispersions, the 
quantities that will play a key role in our study. 

We will use the same techniques as \cite{2005MNRAS.362....9B} to match
the MaNGA galaxies to NVSS and FIRST and determine radio fluxes.
Using the well-measured properties of the MaNGA galaxies will 
allow us to separate AGN and star formation-driven radio 
emission accurately.
We will quantify the radio activity using the Eddington ratio. Here, the Eddington ratio 
refers to the ratio of the radio-based estimate of the bolometric 
luminosity to the black hole mass, the latter estimated using MaNGA's velocity dispersion.  

The Eddington ratio is a somewhat artificial quantity in 
our context, since the radio luminosity does not directly 
trace the radiation emitted near the black hole, and the
structure of  the accretion flow is surely not spherical. 
However, it is one potentially meaningful
way to measure the AGN power relative to the black hole 
mass. 

The measurements we seek are similar to previous results 
on the distribution of radio AGN luminosities based on 
SDSS-I, SDSS-II, FIRST, and NVSS (summarized thoroughly by
\citealt{heckman14a}). Those results showed that the radio
luminosity function is a strong function of stellar mass;
the characteristic radio luminosity has a 
greater-than-linear dependence on stellar mass or 
estimated black hole mass, suggesting
that the Eddington ratio distribution also shifts to higher
Eddington ratios with mass. \citet{heckman14a} found
that the mean ratio of radio luminosity to stellar or black hole 
mass varied relatively little with stellar population
age (as traced by the $D_n$4000 measurement of the 
4000 \AA\ break; bottom right panel of their Figure 14). 
However, the measurement shown in that figure is potentially affected by 
selection effects, because it shows the ratio of two observed
distributions, each affected by different flux
selection limits, and neither of which are corrected for those
limits. Nevertheless, \citet{janssen12a} found a similar result, 
measuring a roughly constant fraction of radio AGN above a fixed 
luminosity, as a  function of star formation rate (using an H$\alpha$ 
based indicator). In any case, these studies comprise what is a 
surprisingly  small amount of  investigation in the literature
regarding the important question of how radio  AGN luminosity 
statistically correlates with global galaxy star formation rate.

Using MaNGA integral field spectroscopy
from SDSS-IV instead of single-fiber spectroscopy 
from SDSS-I and SDSS-II yields a smaller sample but
a number of advantages relative to the previous efforts
just described. We have more precise measurements
of central velocity dispersion and stellar mass via the deeper
spectra. The star formation rates are more precise and can
be measured by adding up the H$\alpha$ emission in only those regions of a galaxy 
whose emission line ratios are consistent with the presence of star formation
(\citealt{Sanchez_2022}). This allows us to avoid confusion between 
``high excitation radio galaxy emission'' (i.e. a simultaneous optical
narrow line AGN and radio AGN) and star formation. As we show, the MaNGA
Pipe3D star formation rates of \cite{Sanchez_2022} also 
allow us to far more reliably
distinguish star formation powered and AGN powered radio 
emission than the \cite{best12a} sample can using $D_n$4000.
Finally, our analysis more explicitly accounts for the 
selection limits in the radio necessitated by the presence of 
star formation related radio emission.

Section \ref{sec:data} describes the data sets we are 
using in detail. Section \ref{sec:methodology} describes
our AGN identification method and our methods for fitting
and assessing models of the Eddington ratio distribution.
In Section \ref{sec:models} we use these methods to fit 
the parameters of the Eddington ratio distribution and 
search for a variation of this distribution as a function
of galaxy properties. In that section we find that the Eddington 
ratio distribution depends directly on stellar mass but not
directly on any of the other quantities we investigate.
We summarize our conclusions in Section \ref{sec:conclusions}.

\section{Data Samples}
\label{sec:data}

\subsection{Source data: MaNGA, NVSS, and FIRST}
In this section we describe the catalogs we use in the study, i.e. MaNGA, the NVSS radio catalog and the FIRST radio catalog. We make use of two different catalogs
based on the MaNGA data, the DRPall file (\citealt{2021AJ....161...52L})
and the Pipe3D Value Added Catalog (\citealt{2016RMxAA..52...21S}, \citealt{2016RMxAA..52..171S}).

The full MaNGA sample of galaxies reduced by the MaNGA Data Reduction 
Pipeline is summarized in the DRPall file. The catalog contains 
$11,273$ observations of low redshift 
galaxies ($z < 0.15$). It consists of three main subsamples---Primary, Secondary and Color 
Enhanced---plus an Ancillary sample. The Primary and Secondary subsamples together 
comprise $\sim 83\%$ of the full catalog. 
These two subsamples are designed to have a uniform distribution in 
absolute magnitude $M_i$ (\citealt{2017AJ....154...86W}), a rough
proxy for the stellar masses of galaxies. The Color Enhanced subsample, on 
the other hand, is designed to target galaxies that are rare in the 
NUV-$i$ versus magnitude plane and to provide better statistics for studying 
them. This sample includes scarce galaxies such as low mass red galaxies 
or high mass blue galaxies.  \cite{2017AJ....154...86W} provides weights to 
correct for the sampling rates as a function of galaxy property. We 
experimented with using these weights in our analysis, but found that doing
so has a small but not significant effect on our final results; in part this 
results because our key analysis regresses against stellar mass and star 
formation rate. Therefore, for simplicity, we do not use these weights in 
the analysis we present here.

Out of the $11,273$ entries in the DRPall file, only $10,261$ correspond to individual galaxies. For these $10,261$ entries, 
we match the MaNGA IFU center to the 
FIRST and NVSS radio catalogs, to investigate potential radio activity. 
FIRST is a $1.4 ~\rm GHz$ radio source catalog, with integrated 
flux densities and positions of over a million sources in the sky. 
It has a flux threshold of 1 mJy, a typical rms of 0.15 mJy, 
a resolution of $5$ arcsec, and subarcsecond positional accuracy,
allowing accurate cross-matching with optical catalogs. However, one drawback of the catalog is that its fluxes for extended sources will be underestimated, being resolved
out in the relatively high resolution interferometric measurements (\citealt{becker95a}).
The lower resolution NVSS $1.4 ~\rm GHz$ radio catalog is designed to
mitigate this issue, with a flux limit of $2.5 ~\rm mJy$ and a 
resolution of $~45~ \rm arcsec$. The combination of NVSS and FIRST 
allows us to reliably match the MaNGA objects to their radio counterparts and
obtain accurate radio fluxes. 

Specifically, we follow the procedure outlined in 
\cite{2005MNRAS.362....9B} to 
identify matches between NVSS, FIRST, and SDSS optical catalogs. 
Although \cite{best12a} published a set of matches to the SDSS Legacy
survey using this procedure complete to 5 mJy, we repeat the analysis
here in order to extend the sample to 2.5 mJy and to include additional
objects in the MaNGA catalog  that did not appear in SDSS Legacy. 
For a full description of the procedure, we refer the reader to 
Section~3 of that paper, but we provide a brief explanation in the 
next subsection.

\subsection{Matching MaNGA to NVSS and FIRST}

\subsubsection{General approach}

Following \cite{2005MNRAS.362....9B}, we perform a spatial match of 
MaNGA galaxies to NVSS, using FIRST to help cross-identify sources and
clarify the nature of potential matches.

Around each MaNGA galaxy we collect all NVSS radio sources within a $3$ arcmin 
radius, which is smaller than the typical separation of NVSS sources ($\sim 9 ~\rm arcmin$) but large enough to encompass extended or multi-component radio 
counterparts of the MaNGA galaxy. We refer to galaxies with multiple NVSS matches 
within this radius as multi-NVSS-component matches and to those with a 
single NVSS match as single-NVSS-component matches. 
The multi-component-NVSS sources may be systems such as double jet 
AGNs without a core component, double jet AGNs with a core component, 
single jet AGNs with a core component, etc. The single-NVSS-component 
matches may be any of the above systems but just unresolved by NVSS at 
their distances. However, due to NVSS's low positional accuracy, the radio sources may or may not be truly coincident with
the MaNGA galaxy and this compels us to invoke FIRST's high positional accuracy.

FIRST's high resolution allows a validation of the association of the 
radio sources with the MaNGA objects. Further matching MaNGA and NVSS 
with FIRST, up to a radius of $30~\rm arcsec$, and by using the rules 
and criteria provided in \cite{2005MNRAS.362....9B}, we carefully 
separate the true matches from the likely fake ones, as explained in 
somewhat more detail below.

Our procedure differs from that of \cite{2005MNRAS.362....9B} in that
we use a flux threshold of 2.5 mJy (roughly 5$\sigma$), instead of their 
choice of  5 mJy (roughly 10$\sigma$).  
According to \cite{2005MNRAS.362....9B}, at $z \sim 0.1$, 5 mJy is the 
value at which the local radio luminosity function begins to be dominated 
by AGN rather than by star forming galaxies. In this paper however, we choose
to maximize the number of radio matches, and use the well
measured SFRs from Pipe3D (see below) to distinguish star formation 
related emission from AGN.

The rules and criteria for different types of sources are briefly provided below. The source types include multi-component-NVSS sources such as $``\rm NVSS ~Doubles",~``\rm NVSS ~Triples"$ and single-component-NVSS sources such as $\rm``NVSS~Singles"$ with one, two, or more FIRST counterparts. Suitable candidates for each source type are chosen by carefully studying the positions and fluxes of nearby NVSS sources of each galaxy. To keep the discussion brief, we choose not to go into the details of how the candidates are chosen (see \citealt{2005MNRAS.362....9B} for more information). Also, since the rules of source classification are similar for the various source types, we only explain the rules for NVSS Doubles in detail. The rules for the other source types are mentioned briefly. 
Readers who wish to carry out similar matching procedures should refer to \cite{2005MNRAS.362....9B} for 
a full description. 

\subsubsection{NVSS Doubles}
\label{ssec:NVSSdoubles}

Candidate doubles could potentially be double-jet systems, double-jet plus core systems, or systems where one or both of the NVSS sources are unassociated with the MaNGA galaxy. To accurately identify which of the above possibilities a given candidate belongs to, we do the following:

\begin{itemize}
    \item To check for a double-jet plus core system, we look for FIRST sources very close to the MaNGA galaxy. If it has a FIRST match within $3 ~\rm arcsec$, we accept the galaxy as a radio source.
    \item To check for a double-jet system or a double-jet plus core system where the core component is resolved out by FIRST, we look at the proximity of the NVSS sources to the optical galaxy and the angle subtended by the directions from the galaxy toward each source. If both the NVSS sources are within $60~\rm arcsec$ of the galaxy and the directions subtend an angle $> 135^{\circ}$, we interpret the two sources as two lobes from oppositely-directed jets, and accept the galaxy as a radio source. 
    \item  To check if both the sources are physically unrelated, we examine their proximities to the galaxy. If both the sources are at separations $> 60~\rm arcsec$, it is likely that they are unrelated to the galaxy. Further, we check for FIRST counterparts at separations $> 15~\rm arcsec$ from the galaxy. Such counterparts are not expected to be seen if the two NVSS sources truly are extended features associated with the galaxy, as FIRST would have resolved them out. If instead, three or fewer such counterparts do exist, and their combined flux is more than $50\%$ the combined flux of the NVSS sources, we reject the galaxy as a radio source.
    \item If only one of the NVSS sources lies within $15~\rm arcsec$ of the galaxy, we categorize it as a candidate single until further examination (see Sections \ref{ssec:1NVSS0FIRST}, \ref{ssec:1NVSS1FIRST}, \ref{ssec:1NVSS2FIRST} and \ref{ssec:1NVSS3FIRST}). 
    \item Finally, if the galaxy does not satisfy any of the above conditions, it is flagged for visual inspection.
    
\end{itemize}
\subsubsection{NVSS Triples and Quads}
\label{ssec:NVSStriquad}

Candidate triples could potentially be double-jet plus core systems, double jet systems with an unrelated source, core-jet systems with an unrelated source, etc. The criteria to accurately identify which of the above systems the galaxy belongs to is similar to that of the NVSS Doubles. Candidate triples with a FIRST match within $3~\rm arcsec$ of the galaxy, we accept as radio sources. We check candidates without a $3~\rm arcsec$ match for pairs of NVSS sources that could possibly be classified as NVSS Doubles, using the criteria in Section \ref{ssec:NVSSdoubles}. Further, we reject galaxies as a radio sources using the same method and arguments as that of the doubles. We flag galaxies that were neither accepted nor rejected for visual inspection.

We 
visually inspected all the candidate quads and candidates with more than four NVSS sources within $3~\rm arcmin$ in order to classify them. We found that only 2 out of 6 such candidates had at least one of the NVSS truly associated with them. 

\subsubsection{NVSS Single; No FIRST}
\label{ssec:1NVSS0FIRST}

We accepted as NVSS singles those candidates with the NVSS source within $10~\rm arcsec$ of the MaNGA galaxy and no FIRST matches within $30~\rm arcsec$. These could possibly be either variable AGN that have become much less active between the measurement time frames of NVSS and FIRST, or extended sources whose flux is resolved out by FIRST.

\subsubsection{NVSS Single; One FIRST}
\label{ssec:1NVSS1FIRST}

Candidate NVSS Singles with one FIRST source within $30~\rm arcsec$ could either be just the core of an AGN or an unresolved core-jet system. We accept a candidate galaxy as an NVSS single if it either has a FIRST counterpart within $3~\rm arcsec$ or a nearby FIRST source that was slightly elongated and pointing towards it. The latter condition allows for the inclusion of core-jet systems unresolved by NVSS.

\subsubsection{NVSS Single; Two FIRST}
\label{ssec:1NVSS2FIRST}

Candidate NVSS Singles with two FIRST sources within $30~\rm arcsec$ could be a core and an unrelated source, a core-jet system or a double-jet system. If one of the FIRST sources lay within $3~ \rm arcsec$, we accepted the galaxy as a radio source. We also accepted candidates as radio sources if the FIRST counterparts were roughly of the same size, flux and subtended an angle $> 135^{\circ}$ at the optical galaxy.  

\subsubsection{NVSS Single; Three FIRST}
\label{ssec:1NVSS3FIRST}

Candidate NVSS Singles with three FIRST sources within $30~\rm arcsec$ could be a core and and two unrelated sources, double-jet system and an unrelated source, double jet plus core system etc. We carried out the following checks out on each candidate before accepting or rejecting it as a radio source:

\begin{itemize}
    \item If the galaxy had a FIRST match within $3 ~\rm arcsec$, we accepted it as a radio source.
    \item We checked each pair of the FIRST sources if they fulfilled the conditions for a FIRST double source (Section \ref{ssec:1NVSS2FIRST}). If the galaxy did fulfill the criteria, we accepted it as a radio source.
    \item If the flux weighted mean position of the three sources lied very close to the galaxy and the farther two sources subtended an angle $>135^{\circ}$ at the closest source, then we accepted the galaxy as a radio source.
\end{itemize}

At the end of this procedure, we find that $1,126$ out of $10,261$ MaNGA 
galaxies are radio sources. For these galaxies, we assign fluxes from the NVSS sources,
summing the component fluxes in the cases of multi-NVSS-component sources. 
We make publicly available this matched catalog, described 
in Appendix \ref{sec:Appendix}.

\subsection{Pipe3D VAC}

For the Eddingtion ratio distribution analysis, we use the Pipe3D
catalog of \cite{Sanchez_2022}.
Pipe3D was successfully run on a subset of 
$10,220$ galaxies from DRPall, corresponding to a set of unique 
objects that correspond to galaxies with redshifts in the 
parent NASA Sloan Atlas (\citealt{blanton11a}).
The Pipe3D analysis pipeline is based on 
the FIT3D fitting tool (\citealt{2016RMxAA..52...21S}). This pipeline 
fits each spaxel's continuum light with a stellar population modeled
as a non-negative linear combination of single stellar populations (SSPs).
It uses this model to continuum subtract the spectra to measure the 
emission lines tracing regions with ionized gas.

For our analysis, we use the integrated stellar mass $M_\star$ and SFR, 
estimated in the catalog. We use the SSP-based estimate of $M_\star$.
For SFR, we work with the $\rm H\alpha$-based estimate. This SFR estimate is 
measured for each galaxy by coadding the $\rm H\alpha$ contributions from those 
spaxels whose emission line ratios are consistent with ionization by star formation. 
Specifically, \cite{Sanchez_2022} choose spaxels that lie in the star-forming region 
of the BPT diagram (\citealt{1981PASP...93....5B}), using the demarcation 
specified by \citet{kewley01a}, and by imposing a 
minimum  $\rm H\alpha$ equivalent width of 3 Angstrom. 

Though the Pipe3D catalog also provides an SSP-based estimate of SFR, we choose to work with the above described $\rm H\alpha$ estimate of SFR. While the two estimates agree well for galaxies with high $\rm H\alpha$ based SFR, there is significant discrepancy in the relationship at low $\rm H\alpha$ based SFR, with the SSP based values usually being relatively overestimated. Nonetheless, we note here that for this study, the SSP based estimate of SFR provides the same result as the $\rm H\alpha$ based SFR. 

\subsection{MaNGA's Data Analysis Pipeline}

Another key quantity in our study is the central 
velocity dispersion of stars ($\sigma_{v}$), which we obtained from
MaNGA's Data Analysis Pipeline (DAP) results. This software package 
estimates astrophysical quantities from the DRPall 
datacubes (\citealt{2021AJ....161...52L}). This DAP includes stellar
velocity dispersion determinations using the penalized pixel-fitting 
method of \citet{2004PASP..116..138C}, using nonnegative linear combinations
of stellar templates. The DAPall table summarises the 
results of the DAP, and among other quantities tabulates 
the DAP stellar velocity dispersion measured within one effective radius 
$\left( R_e\right)$ of the galaxy. which is the measure of velocity 
dispersion we use.

We note that Pipe3D also provides a central velocity dispersion. However,
we found that its values were significantly discrepant relative to the DAP 
values (at the $\sim 50 \%$ level). The relationship between stellar mass
and velocity dispersion has more scatter for the Pipe3D values. A 
consequence of this scatter is that a handful 
of relatively high stellar mass galaxies (5--10 of them) exist with 
radio AGN emission, but with low velocity dispersion in 
Pipe3D. Although these galaxies do not strongly change our overall conclusions,
they do induce a small degree of dependence of AGN fraction on velocity dispersion
when we use the Pipe3D velocity dispersions, which does not appear using the 
DAP values (Section \ref{sec:models}). We interpret these effects to be a result 
of errors in the Pipe3D values, and therefore choose choose to work with the DAP 
values for $\sigma_v$.

\subsection{Final Sample Definition}

To create a sample suitable for our analysis, we start with the 
10,220 galaxies in Pipe3D and we exclude the following:
\begin{itemize}
    \item $2,781$ galaxies with $\sigma_v$ lower than MaNGA's velocity resolution value of $65 \rm ~km~s^{-1}$,
    \item $582$ galaxies with invalid SFRs,
    \item $477$ Ancillary galaxies,
    \item $85$ galaxies that do not have a well-determined redshift,
    \item $13$ galaxies outside the range $-14 \leq \log_{\rm 10}\left({\rm sSFR}/ {\rm y}r^{-1}\right)\leq -9$,
    \item $11$ galaxies that are duplicates of galaxies already existing in the catalog, 
    \item $11$ galaxies with $\log_{10} \left(M_{\star}/M_{\odot}\right) < 8$ or $\log_{10} \left(M_{\star}/M_{\odot}\right) > 12$, and
    \item $5$ galaxies with $\log_{10} \left(\rm SFR/ M_\odot ~yr^{-1}\right) < -5.5$ or $\log_{10} \left(\rm SFR/ M_\odot ~yr^{-1}\right) > 2.5$.
    
\end{itemize}

This process leaves a total of 6,255 galaxies suitable for our analysis of 
the low redshift radio-AGN Eddington ratio distribution. All of these galaxies
were matched to NVSS and FIRST as part of our analysis of the DRPall sample,
and 913 have radio matches.

\begin{figure*}[h]
    \centering
    \includegraphics[width = 0.99\textwidth]{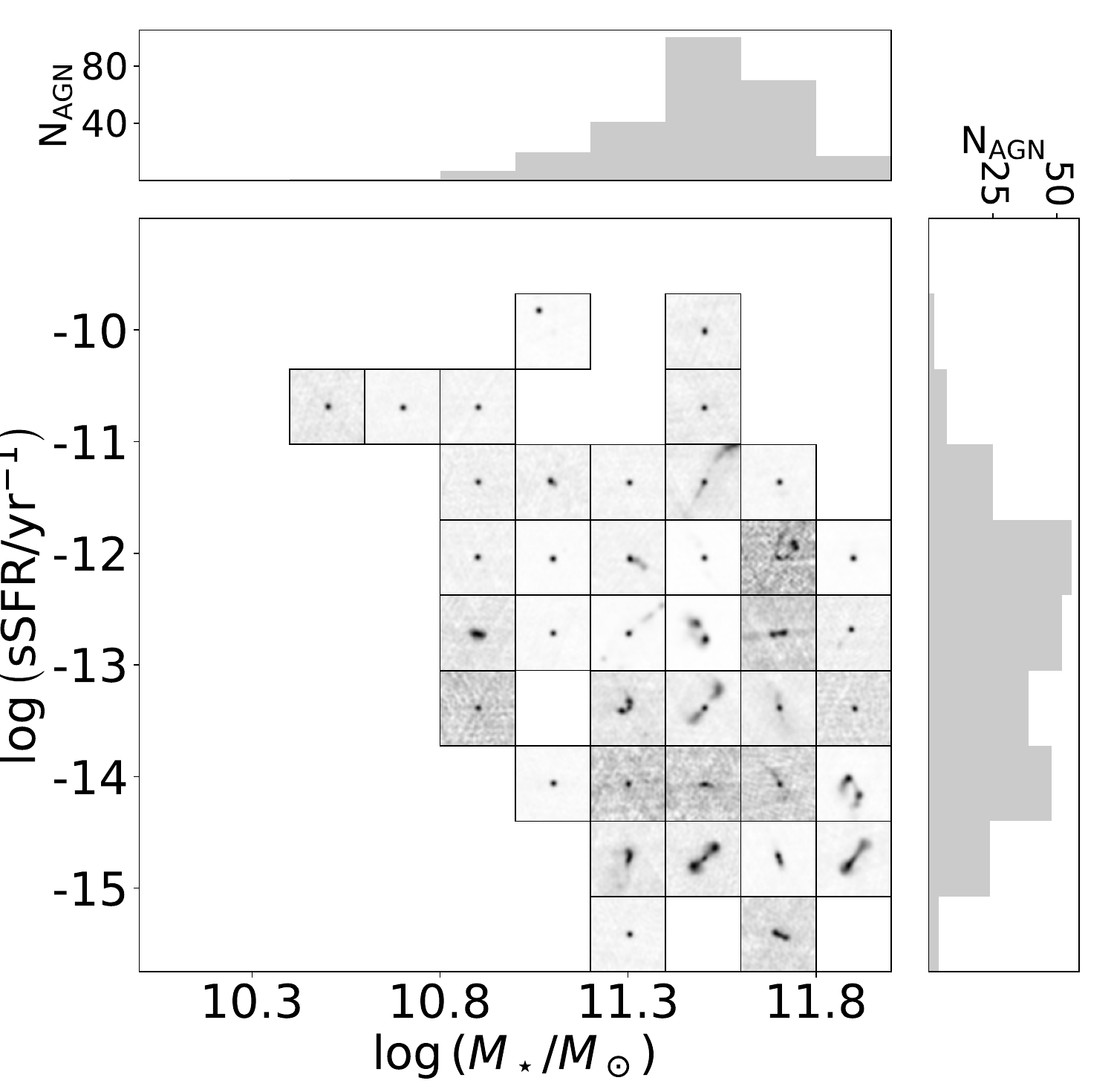}
    \caption{FIRST radio cutouts of detected AGN in bins sSFR-$M_\star$, with each cutout spanning 1.5 arcsec on the sky. In each grid of sSFR-$M_\star$, we have chosen a detected AGN to display. The blank locations in the image are locations where our sample does not have detected AGN. At low $M_\star$ ($\log_{10}\left(M_\star / M_\odot \right) < 10.4$), we see no detected AGN in the sample. At $10.4 < \log_{10}\left(M_\star / M_\odot \right) < 11.3$, the detected AGN mostly seem to be compact sources. And at $\log_{10}\left(M_\star / M_\odot \right) > 11.3$, the AGN population seems to be a mix between compact and extended sources. The purpose of the image is purely for illustration. We do not recommend that any statistical inferences be made based on this image. See Section \ref{sec:conclusions} for more comments on the figure.
    \label{fig:radio_images}}
\end{figure*}

Based on this matching, Figure \ref{fig:radio_images} shows 
$1.5 ~\rm arcmin$ FIRST cutouts of selected radio AGN (see Section \ref{ssec:agnid}), as a function of host galaxy $M_\star$ and sSFR. The marginal histograms show the number of detected AGN in each bin of $M_\star$ and sSFR. Some of the radio images show well defined jets and lobes that are characteristic of radio AGN (\citealt{1974MNRAS.169..395B}, \citealt{1974MNRAS.166..513S}). 
The white panels in the image are bins in which the matched sample does not have any detected AGN. See Section \ref{sec:conclusions} for more detailed comments on the figure.

\section{Methodology}
\label{sec:methodology}

This section describes 
our procedure to distinguish AGN from star formation related emission (Sections 
\ref{ssec:sfr_emission} and \ref{ssec:agnid}),
to estimate Eddington ratios (Section \ref{ssec:ER method}), and to estimate 
the Eddington ratio distribution (Section \ref{ssec:erd_estimate}).

\subsection{Star formation related radio emission}
\label{ssec:sfr_emission}

In addition to AGN activity, 
galactic radio emission can stem from free-free and synchrotron 
processes associated with star formation 
(\citealt{1992ARA&A..30..575C}). Numerous studies have measured
a strong relationship between SFR and radio luminosity (e.g. \citealt{1992ARA&A..30..575C}, 
\citealt{1998ApJ...506L..85C}, \citealt{2003ApJ...586..794B}). 

\begin{figure*}[t]
    \centering
    \includegraphics[width = 0.97\textwidth]{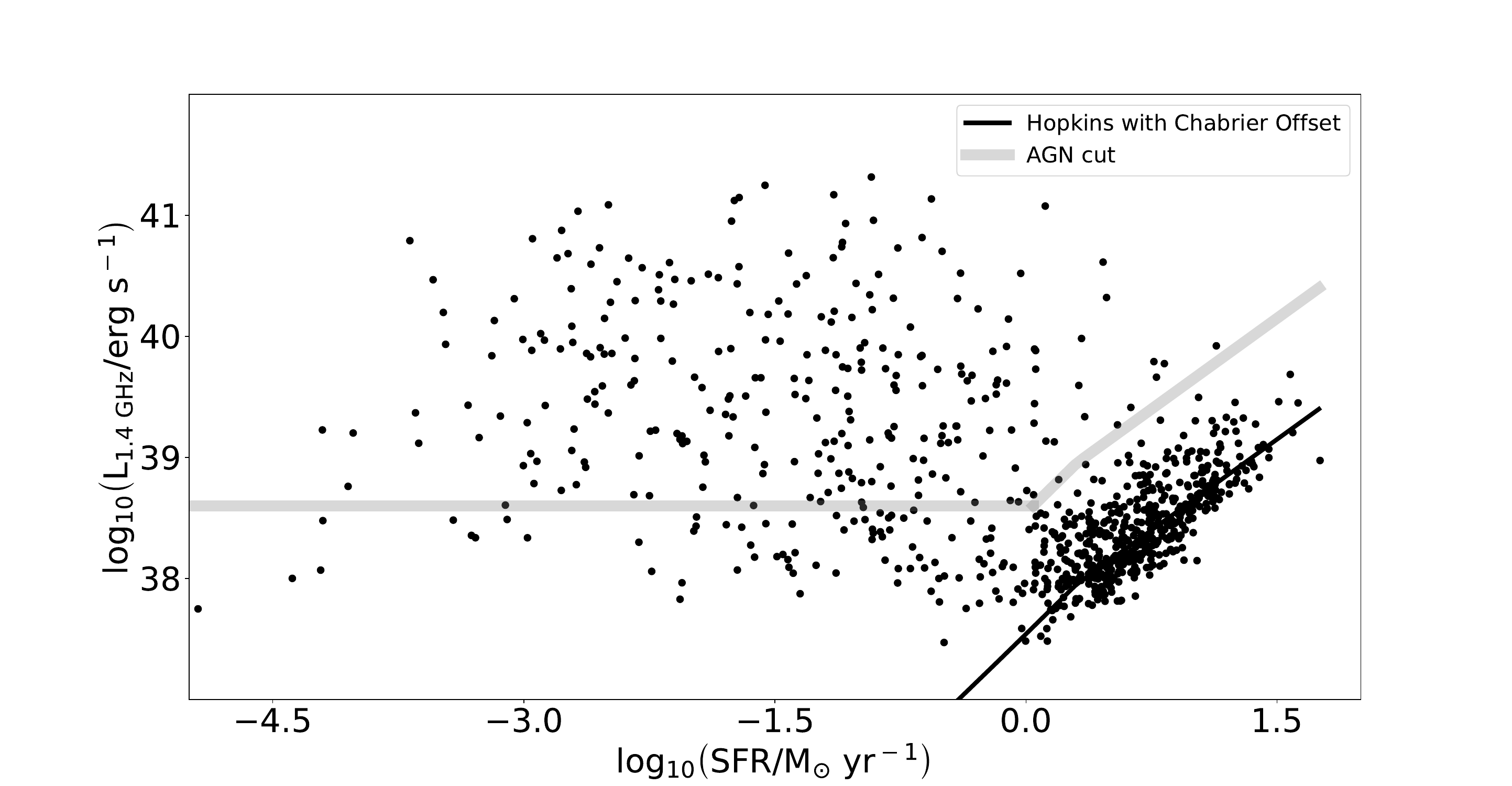}
    \caption{\label{fig:L_vs_sfr} Radio luminosity $\nu L_\nu$ from NVSS at 1.4 GHz
    and $\rm H\alpha$ star formation rates from MaNGA. The solid black line shows the relationship found
    by \cite{2003ApJ...599..971H} between radio synchrotron emission and star
    formation, with an offset to account for the usage of the Chabrier IMF instead of Salpeter. The solid grey line shows the luminosity that 
    must be exceeded for an object to be considered a detected AGN in our analysis.}
\end{figure*}

Figure \ref{fig:L_vs_sfr} shows the MaNGA $\rm Pipe3D$ star formation rates 
and the NVSS $L_{\rm 1.4~GHz} = \nu L_\nu$ at $\nu = 1.4~$GHz for the matching radio detections. 
The objects on the plot fall into two fairly distinct groups. 
The well correlated cloud of points at 
$\rm \log_{\rm 10}\left( SFR/ M_\odot~  yr^{\rm -1}\right) > 0$ 
likely have their radio luminosity dominated by star formation.
MaNGA's high precision SFR 
measurements lead to a reasonably tight relationship between $\rm SFR$ 
and radio luminosity, with a standard deviation of 0.28 dex, 
calculated as described below.
For galaxies with emission greatly exceeding that expected from 
star formation, we ascribe their radio luminosity primarily
to AGN activity.
The radio-star formation relation therefore allows for a reliable and 
well-characterized separation 
of the AGN from star formation related emission.

To compare to previous results, in  Figure \ref{fig:L_vs_sfr} 
we show the $L_{\rm 1.4~GHz}$--SFR relation from \cite{2003ApJ...599..971H}. 
This relation is linear above $\rm \log_{ 
10}\left(L_{\rm c} / erg~s^{\rm -1}\right) = 37.95$, 
and steeper below that luminosity. Expressed as the inferred
SFR as a function of radio luminosity, the relation is
\begin{eqnarray}
\label{eq:lsf}
\log_{\rm 10}\left({\rm SFR}/  M_{\odot} \rm ~yr^{\rm -1}\right) & = &
\log_{10}\left(f\right) - c\cr
& & + \log_{10}\left(L_{\rm SFR}/\rm erg \rm ~s^{-1}\right) 
\end{eqnarray}
where:
\begin{equation}
\label{eq:lsf2}
f = \left\{ 
\begin{array}{cc}
1 &  \mathrm{~if~}L_{\rm SFR} > L_{\rm c} \cr
     \left[0.1 + 0.9 \left(L_{\rm SFR}/ L_{\rm c}\right)^{0.
     3}\right]^{-1} &   \mathrm{~if~} L_{\rm SFR} \leq L_{\rm c}
\end{array} \right.
\end{equation}
We numerically invert this equation to determine the expected 
luminosity given a MaNGA-based SFR.

To compare to the Pipe3D results we convert the SFR estimate of 
\cite{2003ApJ...599..971H} from their \cite{1955ApJ...121..161S} stellar
initial mass function (IMF) assumption to Pipe3D's \cite{2003PASP..115..763C} 
IMF assumption.  Assuming that the SFR determinations are dominated
by the massive end of the stellar mass function, we find that the 
fractional change in SFRs between the two assumptions should be 
$\sim -0.25~\rm dex$, which we have applied to the black solid line
in Figure \ref{fig:L_vs_sfr}.
This corresponds to the choice $c = 37.65$ based on the value $c=37.4$
found by \cite{2003ApJ...599..971H}. This shifted relation does not precisely 
fit the distribution, but since our focus is not on the radio-SFR relation,
we will not perform a detailed investigation of this difference here. 

Using a fit of a Gaussian distribution to the scatter around the mean 
relationship, we find a standard deviation of $\sigma = 0.28$ dex in the 
luminosity given a SFR. 

\subsection{Radio AGN identification}
\label{ssec:agnid}

A radio detection is defined as a detected AGN if it satisfies 
one of the following criteria:
 \begin{itemize}
 \item For a galaxy with $\log_{\rm 10}({\rm SFR}/ { M_{\odot} {\rm ~yr}^{-1}}) < 0$, we require $\log_{10}(L_{\rm 1.4GHz}/ {\rm erg} {\rm ~s}^{-1}) > 38.6$ to be an
 identified AGN. 
 \item For a galaxy with $\log_{\rm 10}({\rm SFR}/ { M_{\odot} {\rm ~yr}^{-1}}) > 0$, we require $\log_{10} L_{\rm 1.4GHz} > \log_{10} L_{\rm SFR} + 1$. Here, $L_{\rm SFR}$ is the star formation associated radio luminosity derived from Equation \ref{eq:lsf}. Notice the $1$ dex allowance that has been provided.
\end{itemize}
This criterion is shown in Figure \ref{fig:L_vs_sfr} as the solid grey line. 
The galaxies above this ``AGN ~cut'' line are included in the likelihood 
analysis in Section \ref{sec:models} as detected AGN. 

The remainder of the sample consists of galaxies without a detected radio 
AGN, either
because they have no radio detection or because their luminosities fall 
below our  criteria. However,  these galaxies may still host AGN that are 
simply too faint for the  $2.5 \rm~ mJy$ flux threshold of NVSS, are outshone 
by the  star formation in the galaxy, or just are not luminous enough to 
exceed the AGN thresholds defined above.  For these 
galaxies, we can determine an upper limit $ L_{\rm lim, AGN}$ on the 
radio luminosity that could be associated with an AGN in that galaxy and 
still not be detected  according to our criteria.

To determine these limits, we first use their redshifts and the NVSS 
flux threshold to determine the radio luminosity upper limit for an 
NVSS detection, $L_{\rm lim}$.  Then the AGN radio luminosity upper limits 
$L_{\rm lim , AGN}$  are defined in the following way:

\begin{itemize}
  \item For galaxies with $\log_{\rm 10}\left({\rm SFR}/ M_\odot~{\rm yr}^{-1}\right) < 0$, if $\log_{\rm 10}\left(L_{\rm lim}/ {\rm erg}~{\rm s}^{-1}\right) < 38.6$, the AGN upper limit $\log_{\rm 10}\left(L_{\rm lim,AGN}/{\rm erg}~{\rm s}^{-1}\right)$ is set to 38.6. Otherwise the limit 
  is just $L_{\rm lim}$.
  \item For galaxies with $\log_{\rm 10}\left({\rm SFR}/ M_\odot~{\rm yr}^{-1}\right) > 0$, if $\log_{\rm 10}\left(L_{\rm SFR}/ {\rm erg}~{\rm s}^{-1}\right) + 1 > \log_{\rm 10}\left(L_{\rm lim}/ {\rm erg}~{\rm s}^{-1}\right) $, the AGN upper limit $\log_{\rm 10}\left(L_{\rm lim,AGN}/\rm {\rm erg}~{\rm s}^{-1}\right)$ is set to $\log_{\rm 10}\left(L_{\rm SFR}/ {\rm erg}~{\rm s}^{-1}\right) + 1$. Otherwise the limit is just $L_{\rm lim}$.
\end{itemize}

Notice the $1$ dex allowance that has been provided in the above definitions. 

\subsection{Eddington ratio measurements}
\label{ssec:ER method}

To determine the Eddington ratio of the AGN, we estimate a bolometric luminosity
from the radio measurements and a black hole mass from the galaxy 
velocity dispersion.
Both quantities are highly uncertain, but for this investigation our 
primary purpose for the Eddington ratio is to study the dependence
of the radio luminosity in a fashion roughly scaled by black hole
mass. The estimated Eddington ratio yields an approximate quantity
for this purpose.

We infer the bolometric luminosity from the radio luminosity following
the method described by \citet{comerford20a}. 
We infer X-ray luminosities using the conversion of \citealt{2015MNRAS.447.1289P}:
\begin{eqnarray}\label{eq:LrLx}
\log_{10} \left(L_X/ {\rm erg}~{\rm s}^{-1}\right) &=& 0.925 \log_{10} \left(L_{1.4 \rm GHz}/ {\rm erg}~{\rm s}^{-1} \right) \cr
&& + 7.1
\end{eqnarray}
We convert this X-ray luminosity  to a bolometric luminosity 
by multiplying by 20, a ratio provided by \citet{comerford20a}. 

Measurements of black hole masses based on motions near galactic centers 
show that black hole mass correlates well with the bulge velocity dispersion 
of the stars (\citealt{gebhardt00a}).
To estimate the black hole mass from this relationship we use the previously described $\sigma_v$ from the DAP and the 
$M$--$\sigma$ determination provided in \citet{2009ApJ...698..198G}:
\begin{equation} \label{eq:msigma}
    \log_{10}\left( M_{\rm BH}/M_\odot\right) = 8.12 + 4.24 \log_{10}\left(\sigma /200 {\rm ~km} {\rm ~s}^{-1}\right)
\end{equation}
We then calculate the Eddington ratio $\lambda = L_{\rm bol} / L_{\rm Edd}$, 
where
$L_{\rm Edd} = 1.26\times 10^{38}\times  M_{\rm BH}/ \rm {M_\odot} {\rm ~ergs~s}^{-1}$.
For all the objects without a detected AGN we can also determine an
upper limit on the Eddington ratio $\lambda_{\rm lim}$ based on $L_{\rm lim ,{\rm AGN}}$ found above.

\subsection{Estimating the Eddington Ratio Distribution}
\label{ssec:erd_estimate}

We estimate the Eddington ratio distribution $\phi(\lambda)$ using a 
maximum likelihood estimator. The data are the set of measured $\lambda$
values and upper limits $\lambda_{\rm lim}$, and the model we use 
for the Eddington ratio distribution incorporates possible dependencies 
on galaxy host properties.

In particular, we choose the Schechter function (\citealt{1976ApJ...203..297S}),
a power law with an exponential cut off. It is defined as:
\begin{equation}\label{eq:schechter}
\phi(\lambda) = \left\{ 
\begin{array}{cc}
\phi_{0} \left(\frac{\lambda}{\lambda_{\ast}}\right)^{-\alpha} \exp\left(-\frac{\lambda}{\lambda_{\ast}}\right) &  \mathrm{~if~}\lambda \ge \lambda_{\rm min} \cr
     0 &   \mathrm{~if~}\lambda < \lambda_{\rm min}
\end{array} \right.
\end{equation}
where $\phi_{0}$ is a normalisation constant,
$\alpha$ is a power law index characterizing the faint end slope,
$\lambda_{\ast}$ is the Eddington ratio characterizing the 
drop-off in the distribution at high values, and $\lambda_{\rm min}$ is 
minimum possible value in the distribution.

We allow this Schechter function to depend on galaxy properties
through a variation of the faint end slope that depends on the
property $Q$ as:
\begin{equation}\label{eq:alpha}
\alpha = \alpha_{0} \left(\frac{Q + k}{Q_0}\right)^\beta,
\end{equation}
where we will take:
\begin{itemize}
\item $Q = \log_{10} \left( M_\star /  M_\odot \right)$, $Q_0 = 10.5$ and $\rm k = 0$ to characterize the mass dependence,
\item $Q = \log_{\rm 10} \left(\sigma_v / {\rm ~km~s}^{-1}\right)$, $Q_0 = 2.2$ and $\rm k = 0$ to characterize the velocity dispersion dependence,
\item $Q = \log_{10}\left( \rm SFR /  M_\odot ~\rm yr^{-1}\right)$, $Q_0 = 4.5$ and $\rm k = 5$ to characterize the star formation rate dependence,
\item $Q = \log_{10}\left( \rm sSFR / ~\rm yr^{-1}\right)$, $Q_0 = 2.2$ and $\rm k = 0$ to characterize the specific star formation rate dependence and,
\item $Q = 1$, $\rm Q_0 = 1$ and $\rm k = 0$ for the case where the model is independent of any galaxy property. We call this the Uniform model.
\end{itemize}
The values of $k$ are set so that the numerator remains positive throughout
the range of the data. $Q_0$ is just a pivot point set near the typical
value of $Q+k$ for the data set. 

This parametric form can be hard to interpret. 
The faint-end slope is not directly constrained very well, and 
$\lambda_{\rm min}$ not directly constrained at all, since it remains 
well below any detectable Eddington ratio.  
The data constrains a nontrivial combination of the faint-end slope,
its dependence on galaxy properties $\beta$, and $\lambda_{\rm min}$.
The power law dependence of $\alpha$ on $Q$ expressed by $\beta$
allows a large range
of different levels of dependence to be modeled, but it is 
not particularly well motivated nor simple to interpret.
We will  see in the results below that a more interpretable 
characterization of the resulting fits is 
the fraction of galaxies $\rm F_{\rm AGN}$ in the model distribution
above some specific $\lambda$ (see Section \ref{ssec:absFrac}).

The likelihood for a single detected AGN becomes:
\begin{equation}\label{eq:likelihood_one}
p_i(\lambda_{i}|\{\alpha_{0},\beta,\lambda_{\ast},\lambda_{\rm min}, Q_i\}) = \rm \phi_{0, i}  
\left(\frac{\lambda_{i}}{\lambda_{\ast}}\right)^{-\alpha_{i}} 
\exp\left(-\frac{\lambda_{i}}{\lambda_{\ast}}\right),
\end{equation}
where $\alpha_i$ is defined according to the galaxy property $Q_i$ and Equation 
\ref{eq:alpha}, and $\rm \phi_{0,i}$ is the appropriate normalization value.

The likelihood for a single AGN non-detection is 
the distribution integrated over the range of possible 
Eddington ratios:
\begin{eqnarray}\label{eq:likelihood_non}
&& p_i(\lambda_{\rm lim}|\{\alpha_{0},\beta,\lambda_{\ast},\lambda_{\rm min}, Q_i\}) = & \cr 
&& \quad \int_{\lambda_{\rm min}}^{\lambda_{\rm lim}}
{\rm d}\lambda\, 
\phi_{0, i} \left(\frac{\lambda}{\lambda_{\ast}}\right)^{-\alpha_{i}} \exp\left(-\frac{\lambda}{\lambda_{\ast}}\right)  
\end{eqnarray}

Over the parameters $\{\alpha_0, \beta, \lambda_\ast, \lambda_{\rm min}\}$, 
we maximize the logarithm of the likelihood:
\begin{equation}\label{eqn:lnL}
\ln \mathcal{L}= \sum_{i = 1}^{N} \ln p_i
\end{equation}
where $N = 6,255$ is the total number of galaxies in our sample, and $p_i$ is calculated
using Equation \ref{eq:likelihood_one} or \ref{eq:likelihood_non}, as appropriate
for each galaxy.
We minimize  $-\ln\mathcal{L}$,  using the Nelder-Mead algorithm
implemented in the {\tt minimize} function in the {\tt optimize} model in SciPy
(\citealt{2020SciPy-NMeth}). 

We furthermore 
obtained model error bars by sampling the posterior functions using {\tt emcee}, 
a Markov  Chain Monte Carlo (MCMC) sampler (\citealt{2013PASP..125..306F}). Here, 
the posterior functions were obtained by multiplying the relevant likelihood 
functions with a constant prior over a reasonable subspace of the model 
parameter space. The MCMC was run with 32 independent walkers, a 
multivariate Gaussian proposal function and a chain length of up to 5000 each. 

As a test of our method, we generate fake data drawn from a Schechter distribution 
and carry out the above described fitting procedure. The procedure was successful in recovering 
within the estimated error bars the true model parameters that were used to generate the 
fake data. 

\section{Results: Eddington ratio distributions}
\label{sec:models}

\subsection{Model Parameters}\label{ssec:modelparms}

\begin{table*}[t]
\centering
\label{tab:results} 
\caption{Schechter function fit parameters to Eddington ratio distributions}
\begin{tabular}{rrrrrr}
\hline
Model & $\alpha_0$ & $\beta$ & 
$\lambda_\ast$ & $\log_{10} \lambda_{min}$ & 
$\ln \mathcal{L}_{i} - \ln \mathcal{L}_{\rm Unif}$ \cr \hline
$M_\star$  & $1.95_{-0.04}^{+0.11}$ & $-4.99_{-0.32}^{+0.27}$ & $0.59_{-0.07}^{+0.24}$ & $-4.68_{-0.17}^{+0.26}$ & 103.76 \cr 
$\sigma_v$ & $1.31_{-0.02}^{+0.11}$ & $-1.69_{-0.20}^{+0.15}$ & $0.55_{-0.04}^{+0.27}$ & $-5.80_{-0.28}^{+0.82}$ & 56.25  \cr 
SFR & $1.44_{-0.04}^{+0.04}$ & $-0.24_{-0.05}^{+0.05}$ & $0.49_{-0.08}^{+0.12}$ & $-4.19_{-0.18}^{+0.12}$ & 12.02  \cr 
sSFR & $1.42_{-0.05}^{+0.03}$ & $0.07_{-0.11}^{+0.07}$ & $0.57_{-0.10}^{+0.17}$ & $-4.23_{-0.26}^{+0.07}$ & 0.02 \cr 
Uniform & $1.42_{-0.05}^{+0.03}$ & 0 & $0.57_{-0.09}^{+0.18}$ & $-4.24_{-0.21}^{+0.08}$ & 0.00 \cr
\hline
\end{tabular}
\end{table*}

Table \ref{tab:results} shows the optimal parameters for each model type $Q$, along 
with their 1 sigma errors. 

The $M_\star$ model has a value of $\beta$ far from zero, suggesting that that the 
Eddington ratio of a radio AGN is heavily dependent on the stellar mass of the 
host galaxy. Galaxies with high $ M_\star$ 
($\log_{10}\left( M_\star/ M_\odot \right) > 10.5$) have a 
significantly greater chance of hosting radio AGNs than the low mass 
galaxies. This result is in agreement with what \cite{2005MNRAS.362...25B} and 
many other investigators find in  their analyses.  

The table also shows the difference between the log likelihood values of the 
$Q$ models and the Uniform model. Each comparison is equivalent to
a likelihood ratio test. In all cases except the sSFR model, the dependence on the 
auxiliary parameter $Q$  is highly favored, with the highest significance 
for $M_\star$. The sSFR model has a value of $\beta$ (and $\ln \mathcal{L}_{i} - \ln \mathcal{L}_{\rm Unif}$) very close to 0, 
suggesting that it is not all that different from the Uniform model. Since the models are identical in  every respect except 
for the  choice of $Q$, these likelihood ratio tests are a meaningful 
way  of comparing the different choices. The table is organised in 
descending order of model likelihood. Clearly, the $ M_\star$ 
model outperforms the other models by a large margin.

All of the four parameters we study are correlated with each 
other to some degree in the MaNGA sample. The superiority of the 
$M_\star$ model relative to the others leads us to ask whether 
the other three parameters matter at all, or if the detected 
dependence on SFR and $\sigma_v$ is just a reflection of how well they 
are correlated with $M_\star$. We will examine this question in the 
following sections, finding that indeed there is little evidence for 
any dependence on the other parameters at fixed $M_\star$. 

\subsection{Monte Carlo Samples}\label{ssec:MC}

To illustrate the results of the model fitting just described, and
to answer the question as to whether there is any dependence of 
the Eddington ratio distribution on parameters other than $M_\star$, 
we will create Monte Carlo samples generated from each of
the five model types, and compare them to the observed
sample.

To create the MC datasets, we perform the following steps. 
For each galaxy in each model, we generate $w = 1000$ Monte Carlo 
values of radio luminosity, summing a radio AGN component and a 
SFR-based component:
\begin{equation}
    L_{\rm 1.4GHz, ~{MC}} =  L_{\rm 1.4GHz, ~{\rm AGN}} + L_{\rm 1.4 GHz, ~{\rm SFR}}
\end{equation}
$L_{\rm 1.4 GHz,~ {\rm AGN}}$ is the radio luminosity corresponding 
to a $\lambda_{\rm AGN}$ that is drawn from the Schechter distribution function,
for the model under consideration and the value of the appropriate
$Q$ for each galaxy (i.e. $M_\star$, $\sigma_v$, SFR, sSFR, or none).
$L_{\rm 1.4 GHz, ~{\rm SFR}}$ is determined from the SFR using the relationship
in Equations \ref{eq:lsf} and \ref{eq:lsf2}, adding Gaussian 
scatter with a standard deviation of 0.28 dex. The luminosity is 
converted to flux using the redshift of the galaxy.

Each galaxy in the MC datasets can be classified as one of three 
possible things, as far as its radio AGN nature is concerned. The 
first classification is as a detected radio AGN under our criteria for 
AGN identification (see Section \ref{ssec:agnid}). 
The second classification is as a detected radio source but not an AGN 
detection under our definition.
The third 
classification is as an undetected radio source given NVSS's flux threshold. 

For each identified radio AGN in the Monte Carlo, we determine 
the Eddington ratio $\lambda_{\rm MC}$ the same way we do for the 
observations, using Equations \ref{eq:LrLx} and \ref{eq:msigma}. 

\begin{figure*}[h]
    \centering
    \includegraphics[width = 7.2in]{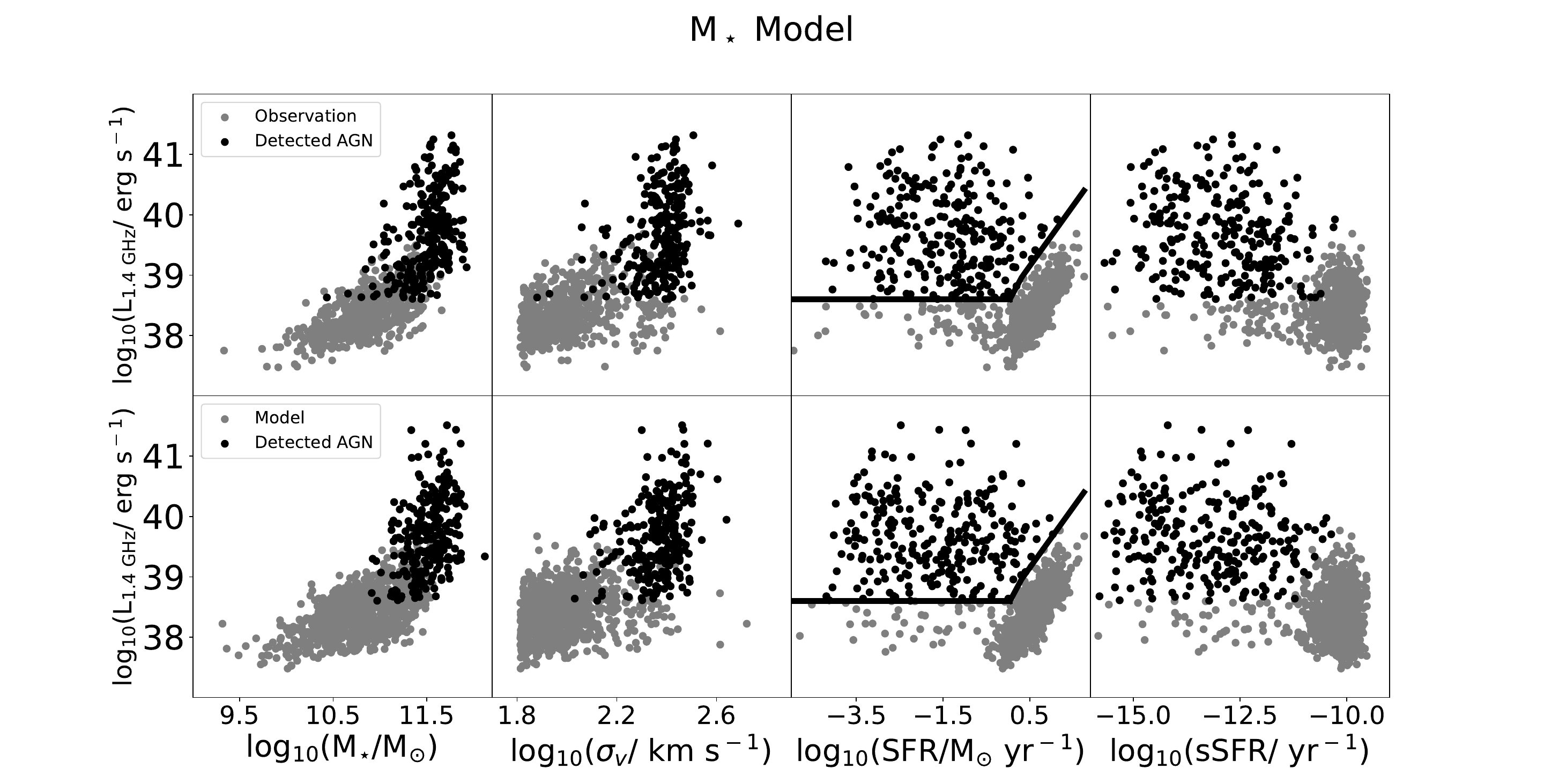}
    \caption{$L_{\rm 1.4GHz}$ versus $M_{\star}$, $\sigma_{v}$, SFR and sSFR
     for the observations (top) and the best fit model
    with $Q=M_\star$ (bottom).}
    \label{fig:L_v_prop(mass)}
\end{figure*}

\begin{figure*}[h]
    \centering
    \includegraphics[width = 7.2in]{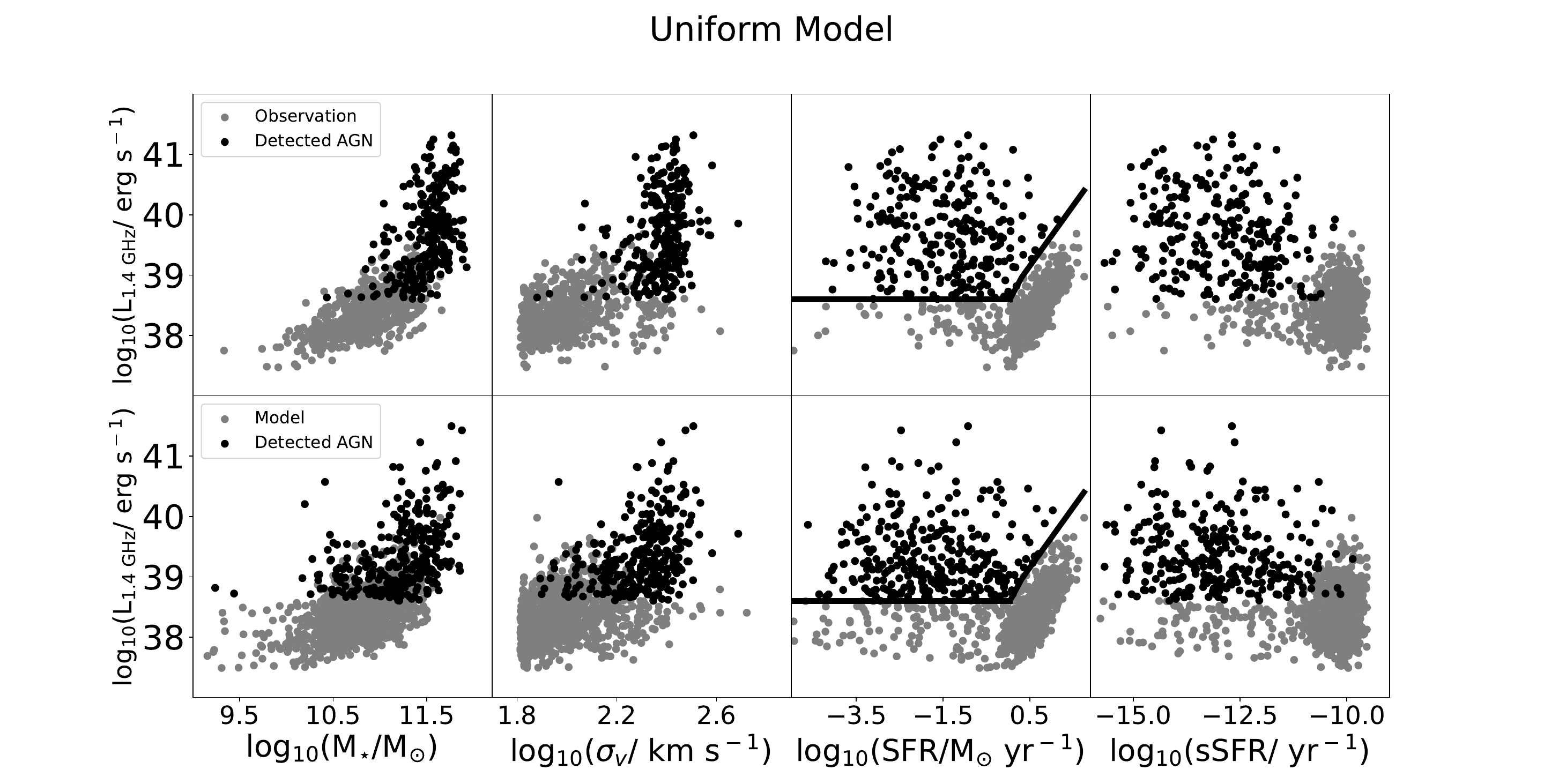}
    \caption{Similar to Figure \ref{fig:L_v_prop(mass)}, with the bottom 
    row corresponding to the best fit Uniform model.}
    \label{fig:L_v_prop(unif)}
\end{figure*}

Figures \ref{fig:L_v_prop(mass)} and \ref{fig:L_v_prop(unif)} each show
one of the 1000 Monte Carlo samples, for the models with $Q=M_\star$ ($M_\star$ model)  and $Q=1$ (Uniform model) respectively.
Figure \ref{fig:L_v_prop(mass)} shows the 
Monte Carlo $M_\star$ model predicted distributions (bottom panel) 
and the corresponding observations (top panel). 
Notice the strong dependence of radio AGN activity on stellar mass 
in the first column of the figure. This strong dependence reflects the large
value of $\beta$. Also notice the third column of the first row of the figure. Its distribution is identical to the one in Figure \ref{fig:L_vs_sfr}. Here however, we choose to show the radio detections in grey and the detected AGN in black. The $M_\star$ model overall looks like it matches the observations well 
in all of the columns. As in the observations, for the model there are 
few detected AGN at high SFR or high sSFR. The distribution of the detected AGN in all the columns of the model tends to closely mimic that of the observations. We will characterize this agreement more quantitatively in the following 
sections.

Figure \ref{fig:L_v_prop(unif)} shows an analogous plot for the Uniform model. While the general trends of the distributions are encapsulated even by this model, it fails to get some of the details right. In particular, it predicts too few high luminosity AGN 
($\log_{\rm 10}\left(L_{\rm 1.4GHz}/\rm ~erg~s^{-1} \right) > 41$) 
relative to the observations. Due to the inability of the model 
to account for the tight dependence of the activity on $M_\star$, its 
distributions contain significantly more scatter than the observations. 
Furthermore, it fails to distribute the detected AGN in all the columns as similarly to the observations as the $M_\star$ model. Particularly notably, at low $M_\star$ or 
$\sigma_v$ there are many more detected AGN in the Uniform model than in the 
observations.

\subsection{Dependence of Eddington Ratio Distribution on Host Galaxy Properties}\label{erdplots}

\begin{figure*}[h]
    \centering
    \includegraphics[width = 7.2in]{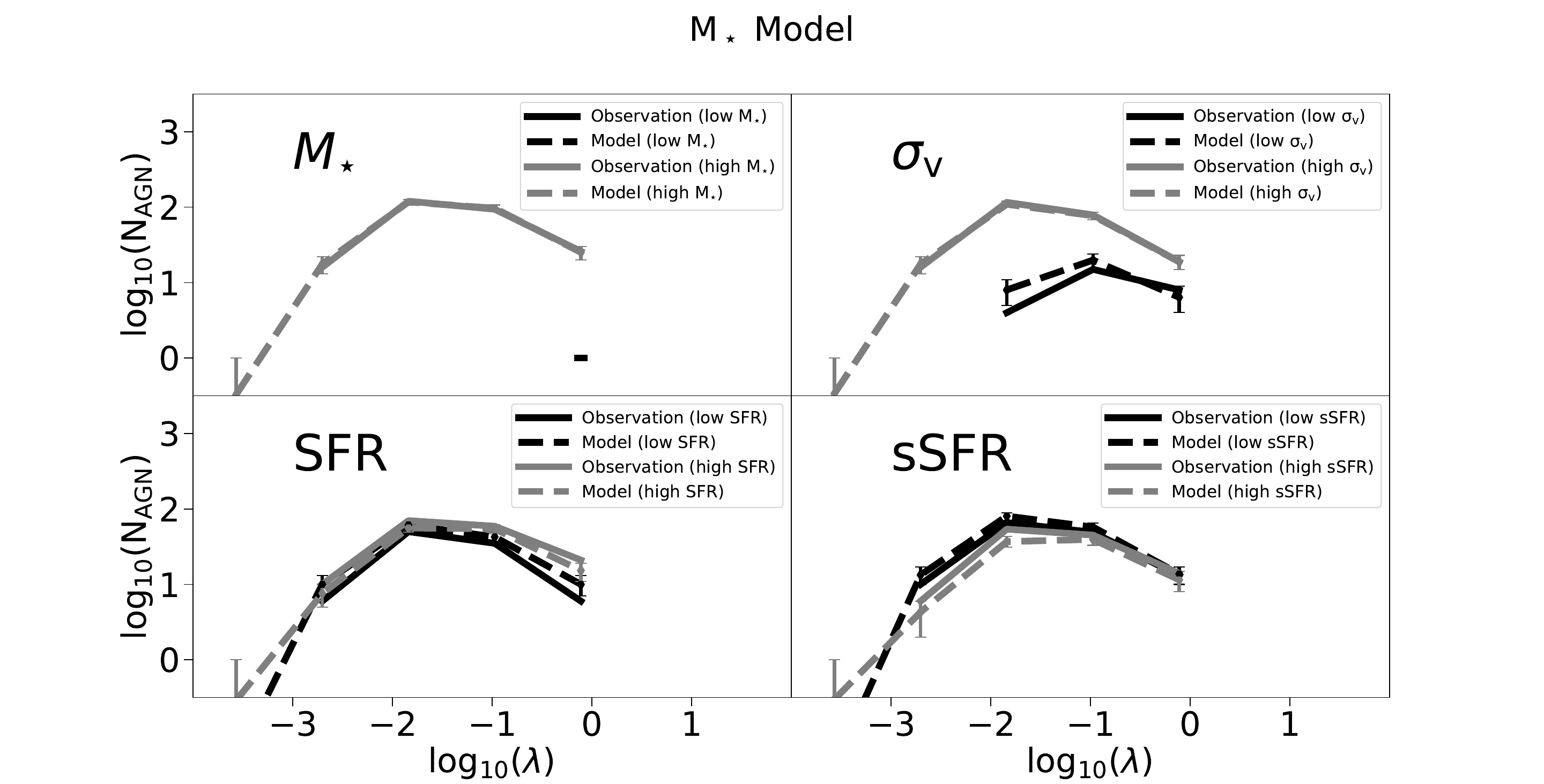}
    \caption{Numbers of AGN as a function of Eddington ratio,
    for two bins (high and low) in each of four properties ($M_\star$, $\sigma_v$, SFR and sSFR). The solid lines connect the observed
    number in each bin. The dashed lines show the model expectation
    value using the $M_\star$ model
    and the  expected 1-$\sigma$ distribution around it.}
    \label{fig:erd(mass)}
\end{figure*}

Figure \ref{fig:erd(mass)} shows the observed and the $M_\star$-model-
predicted Eddington ratio distributions of detected AGNs as functions of $M_\star$, $ \sigma_v$, $\rm SFR$ and $\rm sSFR$. The range of each property has been 
divided into two bins (low and high) and the model is 
compared to the observed data in both of them. 
Table \ref{tab:Qranges} lists the low and high bin ranges for the 
four properties.

\begin{table*}[]
\centering
\caption{Definition of high and low bins for Figures \ref{fig:erd(mass)}---\ref{fig:erd(ssfr)}.}
\label{tab:Qranges}

\begin{tabular}{ccc}
\hline
Property  & Low Bin                                                            & High Bin                                                          \\ \hline
$ M_\star$ & $8.75 < \log_{\rm 10}\left(M_\star/ M_\odot \right) \leq 10.44$    & $10.44 < \log_{\rm 10}\left(M_\star/ M_\odot \right) \leq 12.12$  \\ 
$\sigma_v$ &
  $1.81 < \log_{\rm 10}\left(\sigma_v/~{\rm km ~s}^{\rm -1}\right) \leq 2.27$ &
  $2.27 < \log_{\rm 10}\left(\sigma_v/~{\rm km ~s}^{\rm -1}\right) \leq 2.72$ \\
SFR &
  $-5.28 < \log_{\rm 10}\left(\rm SFR/~M_\odot ~yr^{\rm -1}\right) \le -1.76$ &
  $-1.76  < \log_{\rm 10}\left(\rm SFR/~M_\odot ~yr^{\rm -1}\right) \le 1.76$ \\ 
sSFR      & $-15.85 < \log_{\rm 10}\left({\rm sSFR}/ {\rm yr}^{\rm -1}\right) \leq -12.66$ & $-12.66 < \log_{\rm 10}\left({\rm sSFR}/ {\rm yr}^{\rm -1}\right) \leq -9.53$ \\ 
\hline
\end{tabular}%
\end{table*}

For the low and high bins for each property, Figure \ref{fig:erd(mass)} 
shows the number of detected AGN $N_{\rm AGN}$ in five bins of 
Eddington ratio,  for the observations
and also for the expectation value of the $M_\star$ model. We evaluate the 
expectation value $\bar N_{\rm AGN}$ as the mean of all of the Monte Carlo samples.
We also use the Poisson distribution with a mean $\bar N_{\rm AGN}$ to
determine the expected 68\%
distribution (i.e. ``1-$\sigma$'') around the expectation value. 
This error bar allows us to assess how far away the observations are from
the expectations given the best-fit model. Except for one 
$\lambda$ bin in the low $M_\star$ bin, the contributions of each bin to
the total $\chi^{2}$ are of order unity or less. This evidently means that 
the observations are consistent with being drawn from our best fit model.

In the above mentioned bin, the contribution to $\chi^{2}$ value is $\sim 700$. 
This large value is caused due to one detected AGN with $M_\star \sim 10.43$, 
whose host galaxy has a very high ellipticity of $\sim 0.95$. This can cause the 
SFR estimate of the galaxy to be imprecise, which if the SFR is highly 
underestimated would mean that our classification of it as a detected AGN was 
incorrect. See Section \ref{ssec:agnfrac} for more related comments. 

The top left panel shows that 
the $M_\star$ model successfully predicts that the low mass bins 
should have no detected AGN (except in the above mentioned bin), which is something that the other 
models especially fail to reproduce.

\begin{figure*}[h]
    \centering
    \includegraphics[width = 7.2in]{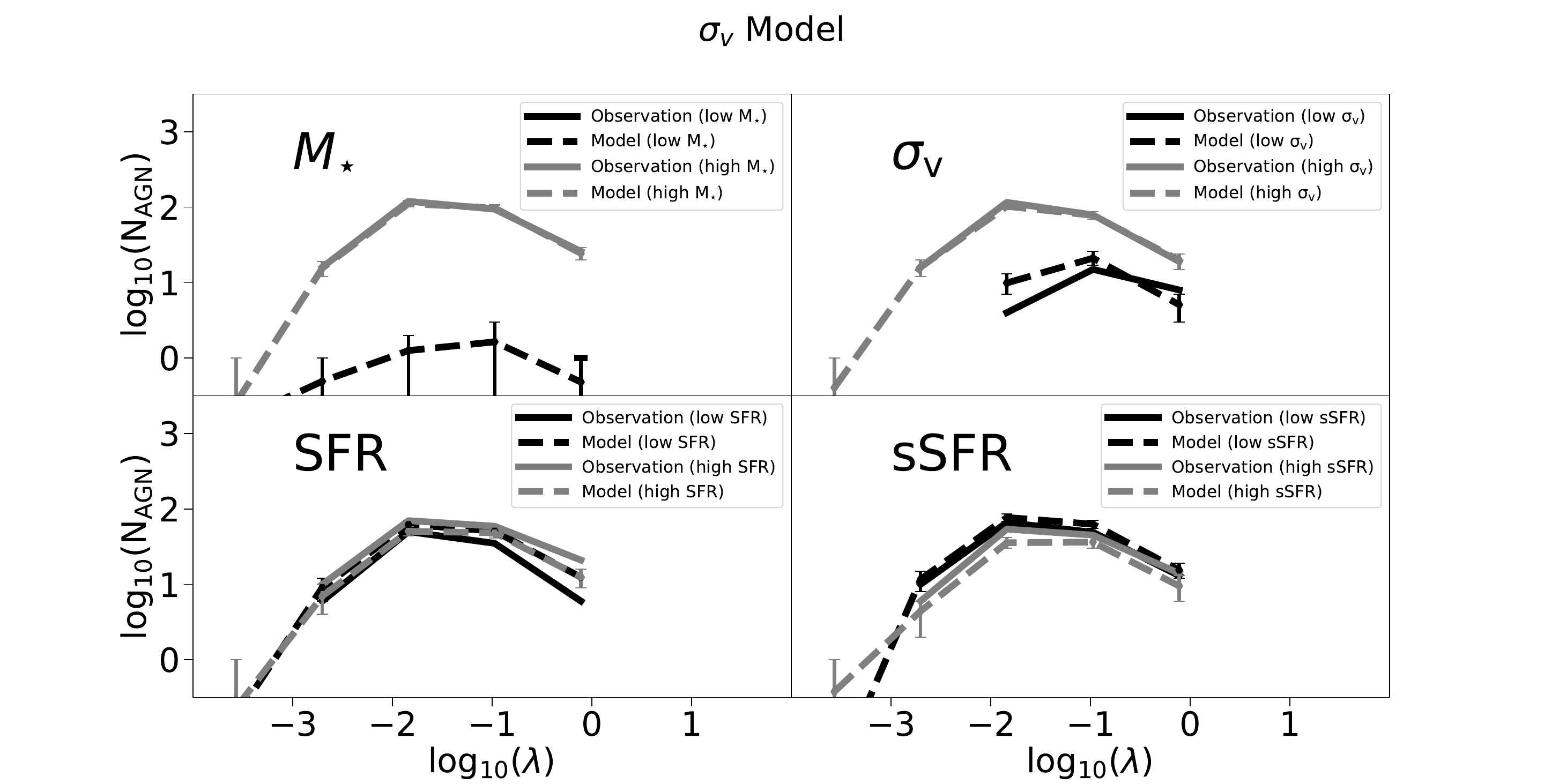}
    \caption{Similar to Figure \ref{fig:erd(mass)}, using the $\sigma_v$ model.}
    \label{fig:erd(sigma)}
\end{figure*}

\begin{figure*}[h]
    \centering
    \includegraphics[width = 7.2in]{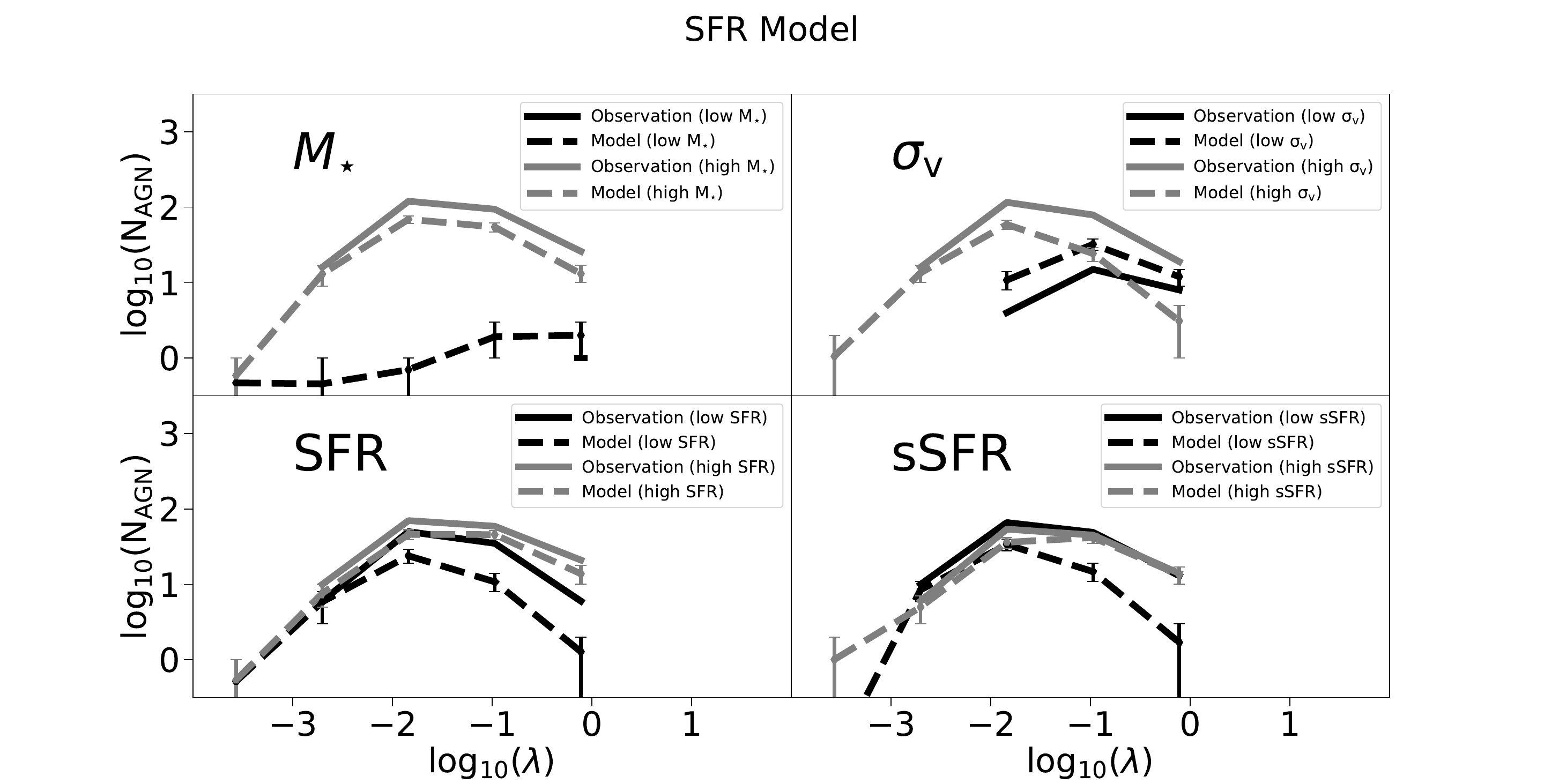}
    \caption{Similar to Figure \ref{fig:erd(mass)}, using the SFR model.}
    \label{fig:erd(sfr)}
\end{figure*}

\begin{figure*}[h]
    \centering
    \includegraphics[width = 7.2in]{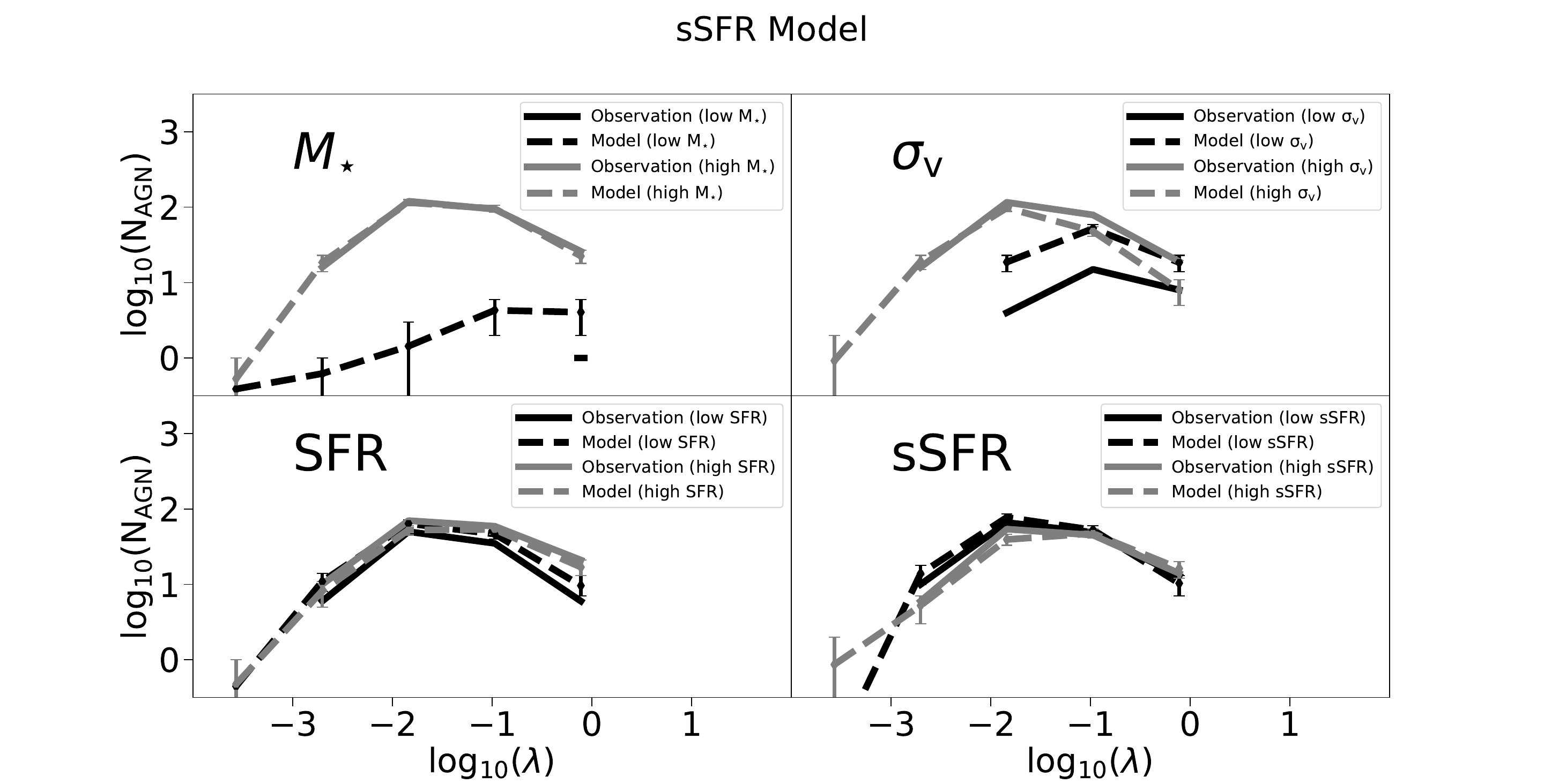}
    \caption{Similar to Figure \ref{fig:erd(mass)}, using the sSFR model.}
    \label{fig:erd(ssfr)}
\end{figure*}

\begin{figure*}[h]
    \centering
    \includegraphics[width = 7.2in]{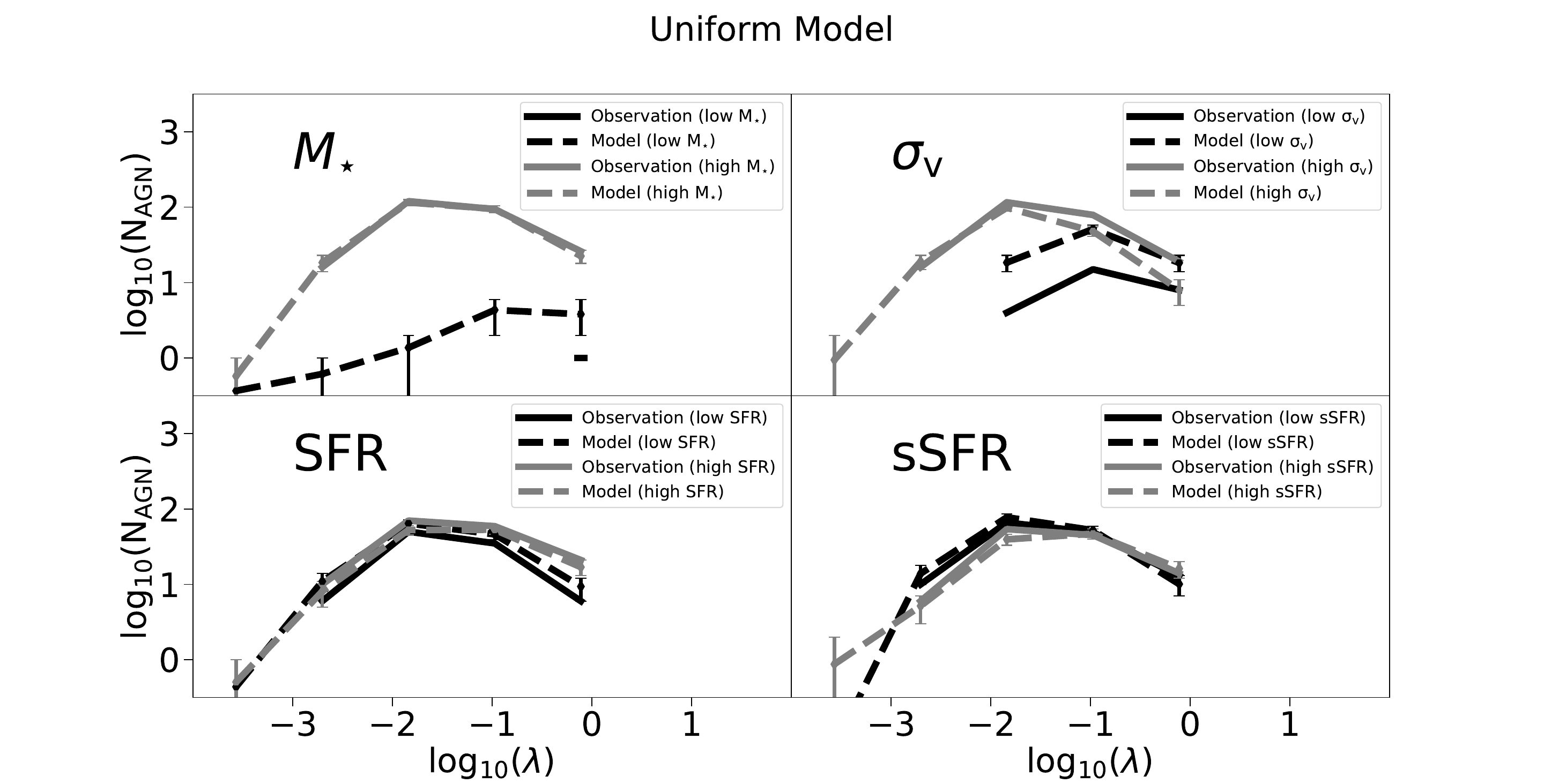}
    \caption{Similar to Figure \ref{fig:erd(mass)}, using the Uniform model.}
    \label{fig:erd(unif)}
\end{figure*}

Figures \ref{fig:erd(sigma)}, \ref{fig:erd(sfr)}, \ref{fig:erd(ssfr)},
and \ref{fig:erd(unif)} show analogous plots for the $ \sigma_v$, $\rm 
SFR$, $\rm sSFR$ and Uniform models, respectively
(see Table \ref{tab:Qranges} for the property bin ranges). As indicated 
by the log likelihood ratios in Table \ref{tab:results}, the 
$\sigma_v$ model comes closest to the $M_\star$ model to 
explaining the observed data. While it manages to do an adequate 
job in most bins, the model predicts too many detected
AGNs in the low $\rm SFR$ and low $ M_\star$ bins.

None of the other models explain the observed Eddington ratio
distributions satisfactorily. While the models follow the 
general trends in the number of observed AGN, the values of the 
observed data do not fall within the estimated uncertainties. 
Neither the $\rm SFR$ nor the $\rm sSFR$ model perform notably 
better than the Uniform model, except with respect to the 
the bins of $\rm SFR$ and $\rm sSFR$ themselves. 

The key observation from these distributions is that the $M_\star$
model shown in Figure \ref{fig:erd(mass)} explains the dependence
of the Eddington ratio distribution on every property. It shows 
no evidence for an additional dependence of the Eddington ratio
distribution on any other of the properties. However, the converse
is not true; when assessing the other models, each one reveals that
there is a required additional dependence on $M_\star$. The natural
interpretation is that the Eddington ratio distribution of the 
AGN is related primarily to the stellar mass $M_\star$, and that 
correlations with other parameters are only due to the secondary
effects of their own correlation with $M_\star$. 

\subsection{Detected AGN Fraction}\label{ssec:agnfrac}

\begin{figure*}[h]
    \centering
    \includegraphics[width = 7.2in]{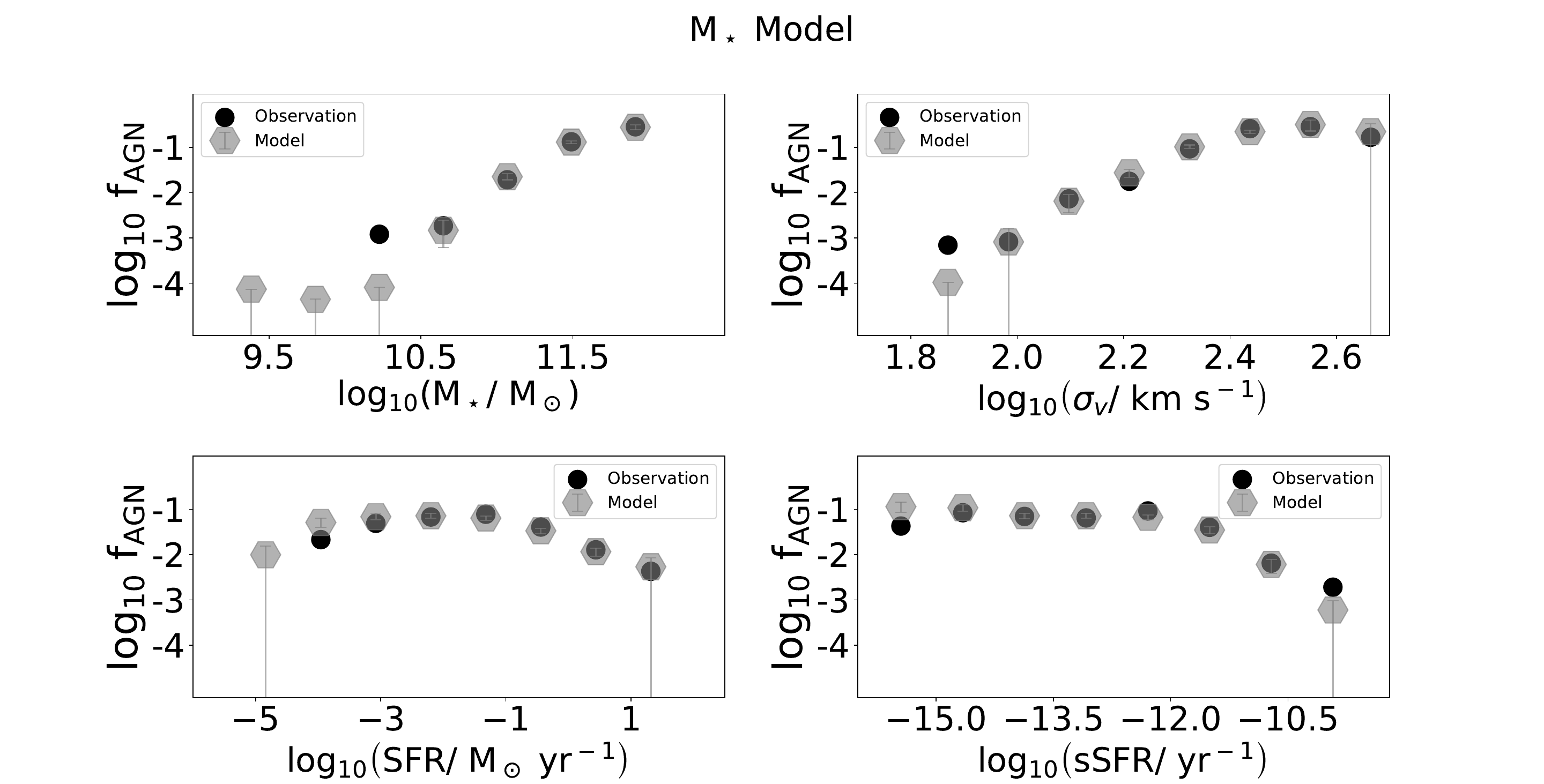}
    \caption{Detected AGN fraction as a function of $M_\star$, $\sigma_v$, SFR and sSFR. 
    The grey hexagons show the expectation value based 
    the best fit model accounting for a dependence on $M_\star$, with
    error bars showing the expected 1-$\sigma$ distribution around the
    expectation value. The black circles show the observations. The
    model, which only accounts for a dependence on $M_\star$, explains
    all of the dependencies of detected fraction as a function of other
    properties.}
    \label{fig:frac(mass)}
\end{figure*}

\begin{figure*}[h]
    \centering
    \includegraphics[width = 7.2in]{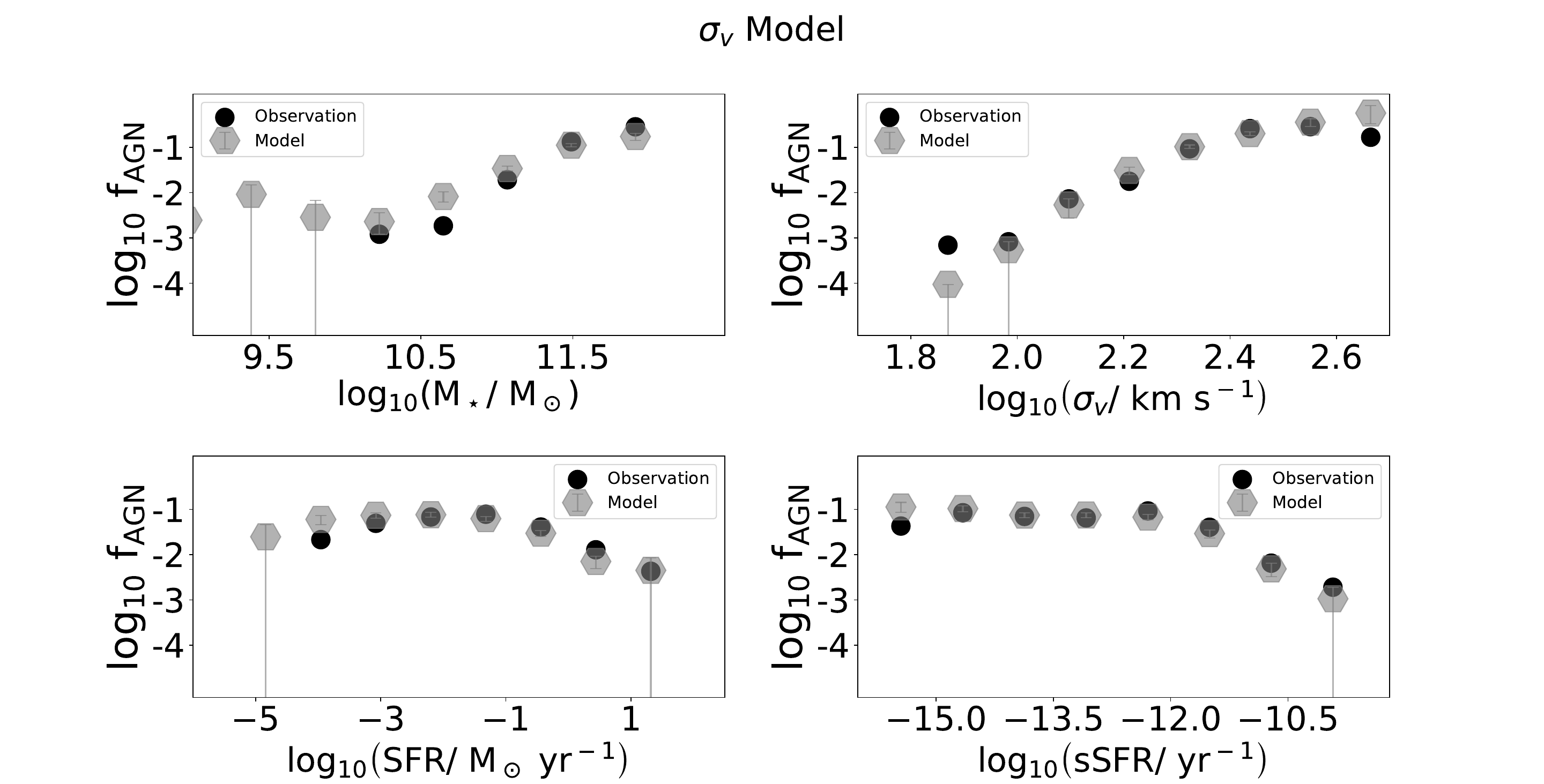}
    \caption{Similar to Figure \ref{fig:frac(mass)}, with identical
    observations but an Eddington ratio distribution model that depends 
    only on $\sigma_v$. This model explains many qualitative dependencies,
    but not as quantitatively well as the $M_\star$ model.}
    \label{fig:frac(sigma)}
\end{figure*}

\begin{figure*}[h]
    \centering
    \includegraphics[width = 7.2in]{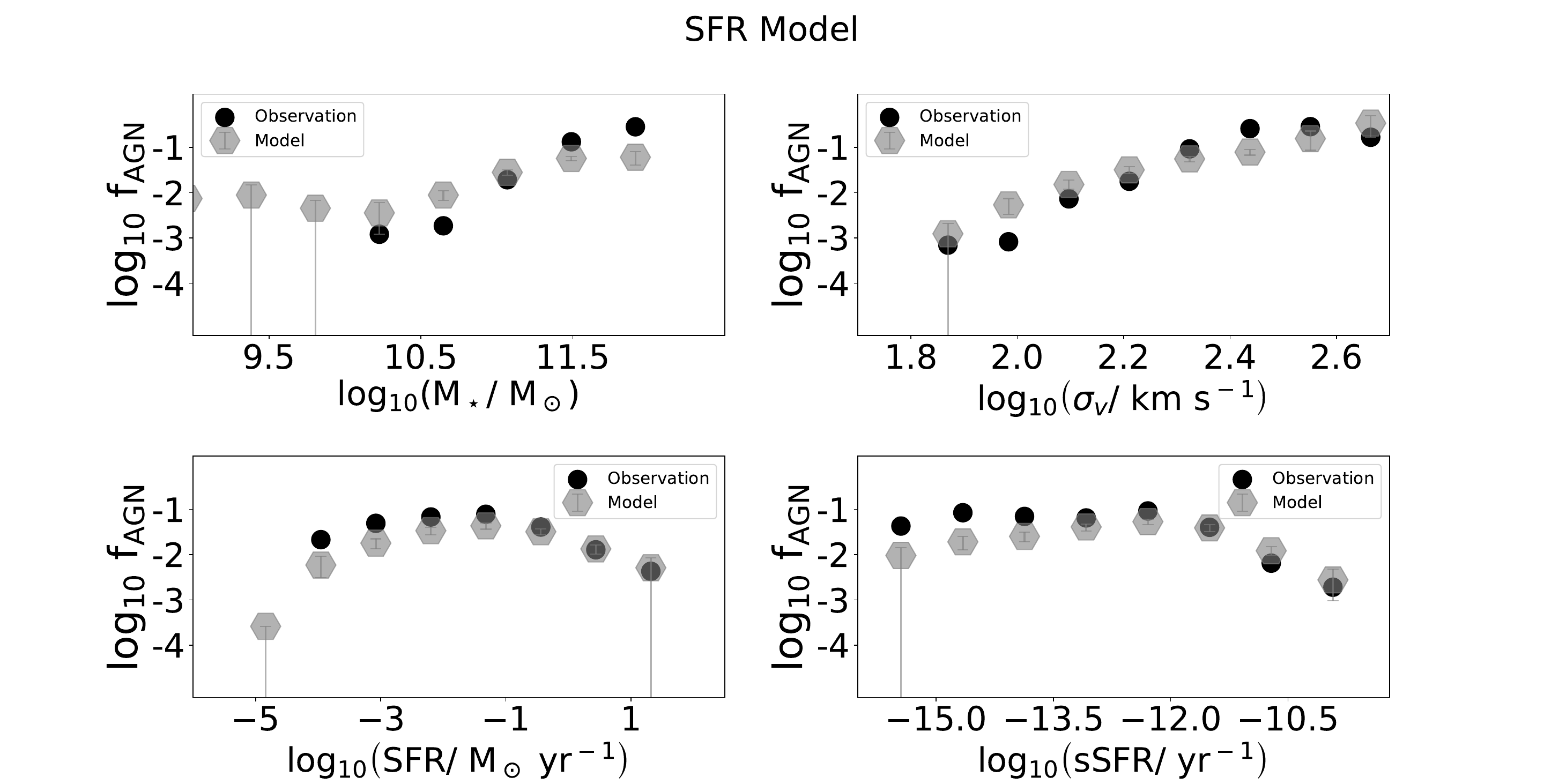}
    \caption{Similar to Figure \ref{fig:frac(mass)}, with identical
    observations but an Eddington ratio distribution model that depends 
    only on SFR. This model does a poor job quantitatively explaining any dependence except
    for that with SFR.}
    \label{fig:frac(sfr)}
\end{figure*}

\begin{figure*}[h]
    \centering
    \includegraphics[width = 7.2in]{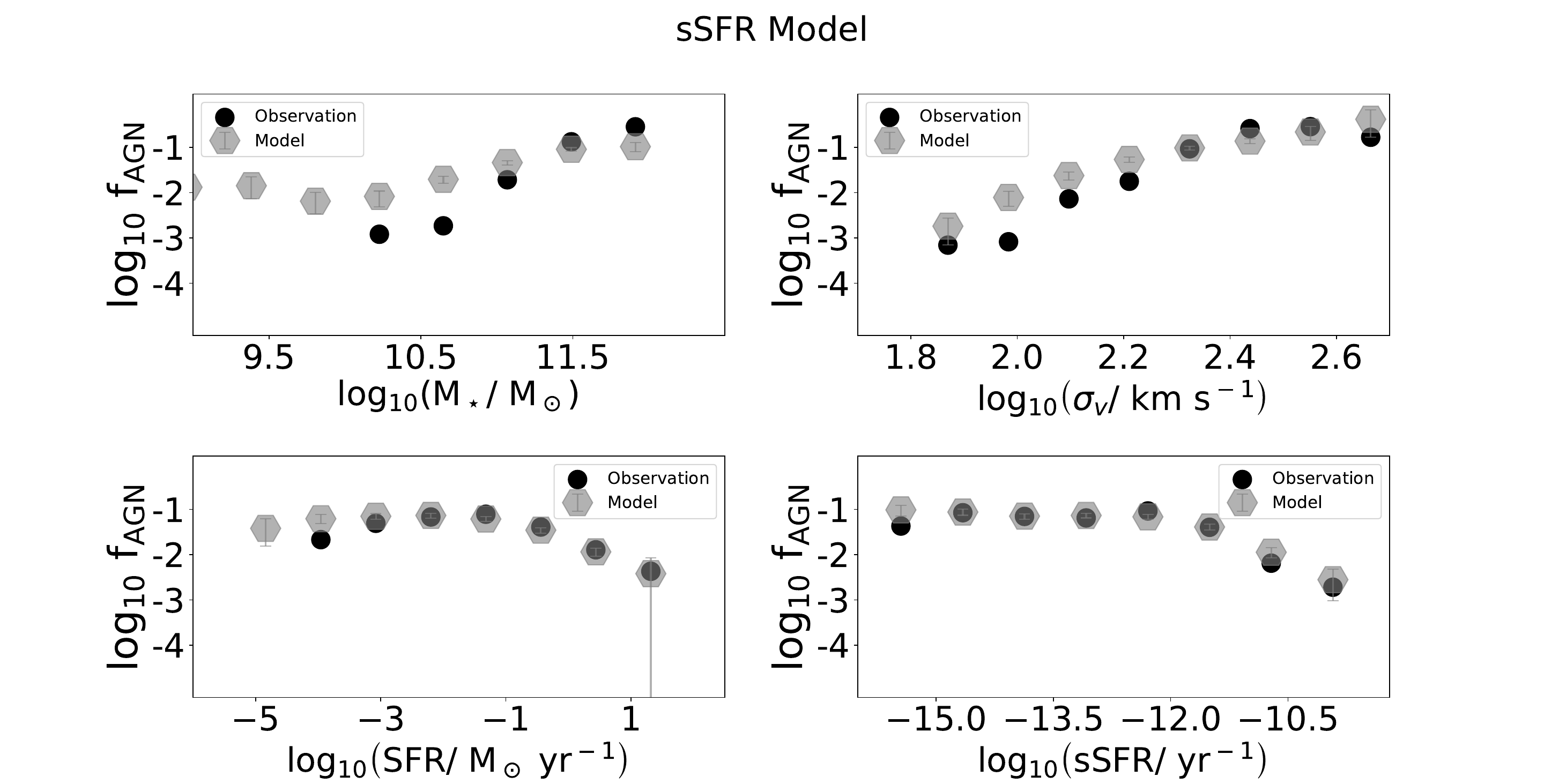}
    \caption{Similar to Figure \ref{fig:frac(mass)}, with identical
    observations but an Eddington ratio distribution model that depends 
    only on sSFR. This model does a poor job quantitatively explaining any dependence except
    for that with sSFR.}
    \label{fig:frac(ssfr)}
\end{figure*}

\begin{figure*}[h]
    \centering
    \includegraphics[width = 7.2in]{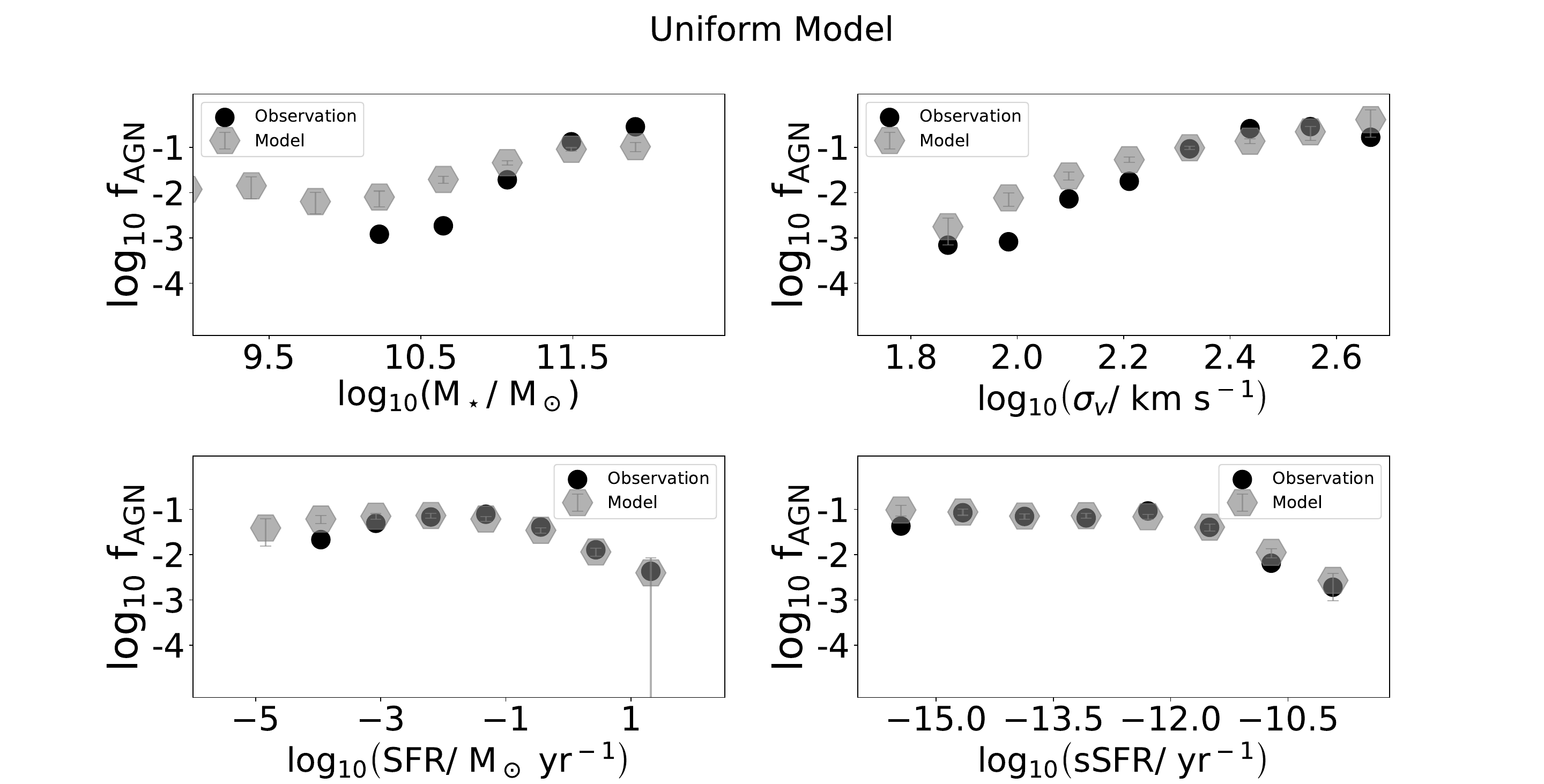}
    \caption{Similar to Figure \ref{fig:frac(mass)}, with identical
    observations but a uniform Eddington ratio distribution model. This model does a poor job quantitatively explaining any 
    dependence, though the qualitative trends are similar to those observed.}
    \label{fig:frac(unif)}
\end{figure*}

In this section, we consider the observed and model-predicted detected
AGN fractions 
in bins of $ M_\star$, $ \sigma_v$, $\rm SFR$ and $\rm sSFR$. 
Figures \ref{fig:frac(mass)}, \ref{fig:frac(sigma)}, \ref{fig:frac(sfr)}, 
\ref{fig:frac(ssfr)} and \ref{fig:frac(unif)} show the detected AGN 
fractions of the $ M_\star$, $ \sigma_v$, $\rm SFR$, $\rm sSFR$ and 
Uniform models against each galaxy property. 

The AGN fraction in a bin is defined as the number of detected AGN in 
the bin divided by the total number of galaxies in the bin. The model 
error bars in each bin correspond to the 1-$\sigma$ limits of the regularised incomplete beta function. 
This beta function is a generalisation of the CDF of the binomial 
distribution to a continuous random variable. 

These plots do a better job at showcasing the dependence of 
AGN demographics on host galaxy properties than the previously shown Eddington ratio
distribution plots. Whereas we divided the ranges in the Eddington ratio
plots (Figures \ref{fig:erd(mass)}---\ref{fig:erd(unif)}) into 
only two bins, we divide them into eight bins here. 

Not surprisingly, the $ M_\star$ model in Figure \ref{fig:frac(mass)} 
outperforms the other models (Figures \ref{fig:frac(sigma)}, \ref{fig:frac(sfr)}, \ref{fig:frac(ssfr)} and \ref{fig:frac(unif)}) in 
predicting the data. Its predictions are almost perfectly consistent with the 
observed trends against all the four galaxy properties. The 
highly negative $\beta$ value of the model leads to a precise reproduction
of the  strong observed trends against $ M_\star$ (see Figure 
\ref{fig:frac(mass)}).

Except in a handful of bins (from the left, the third plotted $M_\star$ bin and
the first $\sigma_v$ bin), 
the contribution towards the $\chi^{2}$ is of order unity or lesser. Though those two bins contribute $\sim 200$ and $\sim30$ respectively to the $\chi^{2}$, they contain only $\sim 1$ detected AGN. 
Furthermore, the AGN in those bins have host galaxies with very high 
ellipticities ($> 0.90$). This can affect their measurement of SFR 
(\citealt{Sanchez_2022}), and hence our radio AGN classification, because our 
definition  of a detected AGN depends on the SFR (see Section \ref{ssec:agnid}). 
Imposing an ellipticity cut on the sample resolves many of the above discrepancies, 
but we choose not to employ that for this study, because we identified this
issue {\it a posteriori}.

Figure  \ref{fig:frac(unif)} shows the fraction plots predicted by the Uniform 
model. Even this uniform model performs reasonably when evaluated against $\rm SFR$ 
and $\rm sSFR$, where it reproduces
the general trends of detected AGN fraction. However, it fails to satisfactorily recreate the observed trends 
against $ M_\star$ and $ \sigma_v$. It is especially bad at predicting 
the correct AGN fractions at low $ M_\star$, reflecting 
results similar to those found in Figure \ref{fig:erd(unif)}, and predicting too
many detected AGN.

The Uniform model in Figure \ref{fig:frac(unif)} is interesting to 
consider in light of what it reveals about how sample selection affects
the properties of the detected AGN sample. This model has the same 
Eddington ratio distribution
for all galaxies. However, the black hole mass correlates with
$\sigma_v$, which itself correlates with the other galaxy properties,
and the AGN detections are further affected by selection effects in 
the sample due to flux limits and our AGN detection limits. These
effects lead to a dependence of the detected AGN fraction on galaxy
properties: a strong rise of detection rates with $\sigma_v$, a slight
rise with $M_\star$, and a decline with SFR or sSFR. These effects 
illustrate why explicitly fitting the Eddington ratio distribution 
accounting for selection effects is so important---as one example,
observing a decline of detected AGN fractions as a function of sSFR 
is not necessarily a sign that high sSFR galaxies have less radio
AGN activity (in an Eddington-normalized sense). Accounting for the 
sample selection is necessary to  draw any accurate conclusions.

The rest of the models reproduce the observed 
trends of detected AGN fractions only with respect to the property 
that the model takes into account. For example,
Figure \ref{fig:frac(ssfr)} 
shows that the sSFR-based model explains the sSFR dependence well, but 
not the other properties. This result reflects similar results to 
those found for the Eddington ratio distribution
in Figure \ref{fig:erd(ssfr)}.

\begin{figure*}[h]
    \centering
    \includegraphics[width = 7.2in]{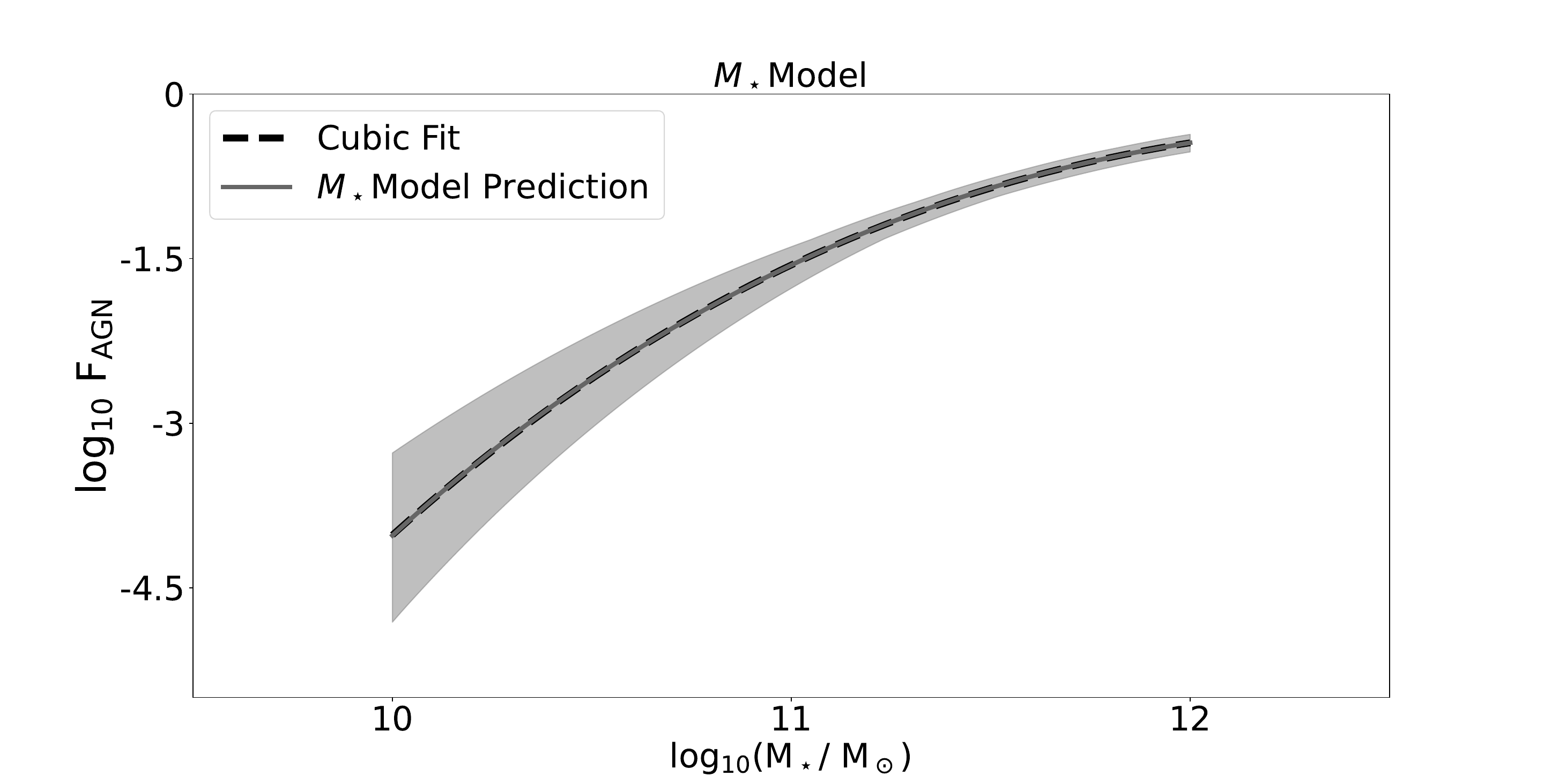}
    \caption{$M_\star$ model predictions for the absolute fraction of AGN with $\lambda>\lambda_c$, where we will take $\lambda_c=0.01$. To the curve, we fit a cubic polynomial in log-log space, as shown by the dashed line. We quote the coefficients of this fit as a specification of our $M_{\star}$ model. Note that this relationship between $F_{\rm AGN}$ and $M_{\star}$ is constrained only in the range $10 \le \log\left(M_{\star}/ M_{\odot}\right) \le 12$.}
    \label{fig:absFrac(mass)}
\end{figure*}

\subsection{Absolute AGN Fraction}\label{ssec:absFrac}

 Our model fits allow us to quantify the level of AGN activity as 
a function of galaxy properties, corrected for the selection effects in the 
sample, by calculating the fraction $F_{\rm AGN}$ of galaxies with 
$\lambda>\lambda_c$, where we will take $\lambda_c=0.01$.
The detected fraction $f_{\rm AGN}$ is affected by the limits $\lambda_{\rm lim}$,
but to the extent that our models are accurate we can simply evaluate them to ask
how $F_{\rm AGN}$ varies as a function of galaxy properties. It would be 
misleading to choose $\lambda_c$ far lower than any $\lambda_{\rm lim}$, and it would
sacrifice statistical power to choose $\lambda_c$ far higher, so the sensible
choice is a $\lambda_c$ similar to the typical $\lambda_{\rm lim}$ values. 
This quantification of the model is more easily interpretable than the 
model parameters themselves.

Figure \ref{fig:absFrac(mass)} shows the $M_\star$ model predicted $F_{\rm AGN}$ plotted against $M_\star$. The error bars correspond to the $68 \%$ limits of the possible values of $F_{\rm AGN}$ as predicted by different 
sets of the model parameters [$\alpha_{0}$, $\beta$, $\lambda_\ast$ and 
$\rm log_{10} \lambda_{\rm min}$] that are drawn from the posterior distribution 
of the MCMC analysis. The general trend of the plot resembles the trend in the top left panel of Figure 
\ref{fig:frac(mass)}, but here are completeness-corrected to account for the 
cases where $\lambda_{\rm lim}>\lambda_c$.

We characterize the mass dependence of $F_{\rm AGN}$ in the model by 
fitting a function to Figure \ref{fig:absFrac(mass)}. This fit is valid in the range $10 \le \log\left(M_{\star}/M_{\odot}\right) \le 12$ and has the following the form:

\begin{eqnarray}\label{eq:Fagn}
&& \log\left(F_{\rm AGN}\right) = a_3(\log\left(M_{\star}\right) - 10.5)^{3} & \cr 
&& + a_2(\log\left(M_{\star}\right) - 10.5)^{2} &\cr
&& + a_1(\log\left(M_{\star}\right) - 10.5) + a_0
\end{eqnarray}

finding $a_3 = 0.07_{-0.03}^{+0.03}$, $a_2 = -0.77_{-0.25}^{+0.22}$, $a_1 = 2.44_{-0.57}^{+0.63}$ and $a_0 = -2.60_{-0.45}^{+0.41}$.

\section{Conclusions}
\label{sec:conclusions}

We have examined the relationship between the Eddington ratio distribution
of radio AGN activity and 
galaxy properties for the MaNGA sample matched with the NVSS and FIRST 
radio surveys. Our results are as follows:
\begin{itemize}
\item Explaining the observations requires that the Eddington ratio
distribution be a strong function of galaxy stellar mass $M_\star$. Figure \ref{fig:absFrac(mass)} shows the required 
dependence in terms of the fraction of AGN with $\lambda > 0.01$ in the 
model fit.
\item Once the stellar mass dependence is accounted for, there is no
evidence for any additional dependence on $\sigma_v$, SFR, or sSFR. 
The agreement of the model with observations in Figures \ref{fig:L_v_prop(mass)},
\ref{fig:erd(mass)}, and \ref{fig:frac(mass)} (and the lack of such agreement
for the other models) demonstrates this fact. 

\end{itemize}

The degree of dependence of the detected AGN fraction on other
properties is fairly stringently constrained.
The variation of the detected AGN fraction across the mass range studied 
is greater than 2 dex. However, the differences between the observed dependence
of the detected AGN fraction on other properties, and that 
predicted by a model that only depends on stellar mass $M_\star$, has an RMS of 0.23 dex (Figure \ref{fig:frac(mass)}). 
The strong dependence of the Eddington ratio distribution
on stellar mass is consistent with many previous studies, 
as summarized by \citet{heckman14a}. But the taxonomy of 
AGN in that paper and many others describes the radio-emitting AGN
as low sSFR galaxies; to 
the extent that such a statement is true, it is almost entirely due to the 
correlation between sSFR and stellar mass, 
because there is little to no independent correlation of sSFR with 
radio AGN activity at a given stellar mass. 

Surprisingly few studies have directly tested this proposition.
As mentioned in the introduction, the lower right panel of 
Figure 14 in \cite{heckman14a} shows that the mean ratio of 
radio luminosity to black hole mass varies little with 
stellar population age; but in that analysis, the ratio shown
has a numerator and denominator affected by different flux
limits that are not accounted for, which can affect the
results. \cite{janssen12a} 
do examine the radio AGN fraction as a function of SFR 
at fixed stellar mass and in their Figure 3
find it nearly constant for
``low-excitation'' radio galaxies (the majority population
that are not narrow line AGN); their figure uses radio luminosities
$\nu L_\nu> 10^{40}$ erg~s$^{-1}$,
so is generally independent of worries about star formation related 
radio emission (cf. our Figure \ref{fig:L_vs_sfr}). That paper emphasizes the 
color dependence of AGN fraction, which is significant at
fixed mass; why this dependence exists 
when the SFR dependence is weak is not entirely clear. 
The work we describe here improves on these previous studies
with superior determinations of star formation rate and a cleaner
treatment of selection effects.

Although our finding is that the Eddington ratio distribution 
is independent of host galaxy SFR, other features of the AGN, such as
the radio morphology, might also be.
We do not directly quantify the dependence of the radio morphology of 
AGN on galaxy host properties, but Figure \ref{fig:radio_images} 
at least does not show any obvious relationship between sSFR and morphology.
It does show an obvious dependence of radio AGN activity and 
morphology with respect to mass. At low 
$M_\star$ ($\log_{10}\left(M_\star / M_\odot \right) < 10.4$), there
are no detected AGN in the sample. At 
$10.4 < \log_{10}\left(M_\star / M_\odot \right) < 11.3$, the detected 
AGN mostly seem to be compact sources. At $\log_{10}\left(M_\star / 
M_\odot \right) > 11.3$, the AGN population seems to be a mix of 
compact and extended sources. The image also shows the lack of 
such an obvious trend with respect to sSFR. Except in the top two 
sSFR bins (where there are very little detected AGN), 
at a given mass, the morphology of the AGN does not change 
substantially with sSFR. That is, the strong trend of radio AGN 
morphology with respect to mass seems to exist irrespective of 
the sSFR of the host galaxy. That being said, we make no claim 
about the relationship between radio-AGN morphology and host 
galaxy properties in this study. The quantification of such a 
relationship requires accurate classifications of all of the radio 
sources and a more detailed statistical investigation, and we 
suggest this as an important topic for future study. 

Radio AGN are assumed in galaxy formation theories to regulate
star formation rates of massive galaxies, and in particular to keep
them quenched, as described in the 
introduction and reviewed by \citet{somerville15a}. The
lack of any relationship between radio AGN activity (in
an Eddington ratio sense) and star 
formation rate may therefore be significant and provides 
an interesting potential test of these theories. The relevant
theoretical predictions, as  far as we know, have not been 
published.

A key aspect of the AGN population is that it manifests in
a wide array of fashions across a large dynamic range of 
wavelengths. They are identifiable in X-rays, UV, optical lines,
and the mid-IR, as well as the radio.
Yet all of these manifestations are subject to the 
selection effects analogous to those described here.
Despite its relatively small size, with its high quality 
measurements of galaxy properties, MaNGA sample lends itself
to the study of AGN in all their manifestations for very well-understood
galaxies, for which characterizing these selection effects is possible.
Along with this
paper we are publishing the catalog of radio AGN detections and upper 
limits. 
We plan to use this catalog in conjunction with 
detections and limits determined in other signatures in
the optical, IR, and X-rays in order to understand
the population as a coherent whole.

\begin{acknowledgements}

Funding for the Sloan Digital Sky 
Survey IV has been provided by the 
Alfred P. Sloan Foundation, the U.S. 
Department of Energy Office of 
Science, and the Participating 
Institutions. 

SDSS-IV acknowledges support and 
resources from the Center for High 
Performance Computing  at the 
University of Utah. The SDSS 
website is www.sdss4.org.

SDSS-IV is managed by the 
Astrophysical Research Consortium 
for the Participating Institutions 
of the SDSS Collaboration including 
the Brazilian Participation Group, 
the Carnegie Institution for Science, 
Carnegie Mellon University, Center for 
Astrophysics | Harvard \& 
Smithsonian, the Chilean Participation 
Group, the French Participation Group, 
Instituto de Astrof\'isica de 
Canarias, The Johns Hopkins 
University, Kavli Institute for the 
Physics and Mathematics of the 
Universe (IPMU) / University of 
Tokyo, the Korean Participation Group, 
Lawrence Berkeley National Laboratory, 
Leibniz Institut f\"ur Astrophysik 
Potsdam (AIP),  Max-Planck-Institut 
f\"ur Astronomie (MPIA Heidelberg), 
Max-Planck-Institut f\"ur 
Astrophysik (MPA Garching), 
Max-Planck-Institut f\"ur 
Extraterrestrische Physik (MPE), 
National Astronomical Observatories of 
China, New Mexico State University, 
New York University, University of 
Notre Dame, Observat\'ario 
Nacional / MCTI, The Ohio State 
University, Pennsylvania State 
University, Shanghai 
Astronomical Observatory, United 
Kingdom Participation Group, 
Universidad Nacional Aut\'onoma 
de M\'exico, University of Arizona, 
University of Colorado Boulder, 
University of Oxford, University of 
Portsmouth, University of Utah, 
University of Virginia, University 
of Washington, University of 
Wisconsin, Vanderbilt University, 
and Yale University.

We thank the members of the Hogg- Blanton research group of New York University, Joseph Gelfand and Ingyin Zaw of New York University Abu Dhabi and the BHM-Pop group of SDSS-V, for their valuable insights that have helped shape this study.
\end{acknowledgements}

\bibliography{refs.bib}

\appendix

\section{Appendix}
\label{sec:Appendix}

Here we describe the MaNGA-FIRST-NVSS value added catalog (VAC) that is provided along with this paper. This catalog includes $10,261$ galaxies in the DRPall file (see Section \ref{sec:data}). To this, we match the NVSS and FIRST radio catalogs, down to NVSS's flux limit of 2.5 mJy. See Section \ref{sec:methodology} and \cite{2005MNRAS.362....9B} for an full explanation of the matching procedure. The columns of the VAC are briefly explained below:

\begin{itemize}

    \item {\tt mangaid}: Unique ID denoting a particular MaNGA object. 
    \item {\tt plateifu}: Plate and IFU bundle numbers of the measured object, separated by a hyphen.
    \item {\tt ra}: Right Ascension of the object (J2000). Units - degree.
    \item {\tt dec}: Declination of the object (J2000). Units - degree.
    \item {\tt fint}: Total integrated radio flux associated with the galaxy, as measured by NVSS. A dummy value of -999 has been provided for galaxies without radio detections. Units - mJy.
    \item {\tt fint\textunderscore error}: Error in the total integrated radio flux associated with the galaxy, as computed using the NVSS data model (\citealt{condon98a}). For multi-component-NVSS matches, the error is the root of the sum of the squares of the errors of individual sources. A dummy value of -999 has been provided for galaxies without radio detections. Units - mJy.
    \item {\tt log\textunderscore lr}: Total radio luminosity of the galaxy after matching with NVSS and FIRST. A dummy value of -999 has been provided for galaxies without radio detections. Units - $\rm \log \left(ergs/ s\right)$.
    \item {\tt radio\textunderscore class}: Following the convention in \cite{2005MNRAS.362....9B}, we categorise the radio sources as one of five classes. Radio Class 1 is a galaxy that has one NVSS and one FIRST match. Radio Class 2 is a galaxy with one NVSS match but multiple FIRST matches. Radio Class 3 is a galaxy with one NVSS match and no FIRST matches. Radio Class 4 is a galaxy with multiple NVSS matches. Radio Class 0 is a galaxy that isn't a radio detection.
    \item {\tt num\textunderscore nvss\textunderscore 3m}: Number of NVSS matches within a $3 ~\rm arcmin$ radius of the galaxy. 
    \item {\tt ras\textunderscore nvss\textunderscore 3m}: Right Ascensions of the $3 ~\rm arcmin$ NVSS matches. Each object in this column is an array of length six. For the galaxies with less than six $3 ~\rm arcmin$ NVSS matches, dummy values of -999 have been provided. The arrays are sorted in descending order of the $3 ~\rm arcmin$ NVSS matches fluxes. Units - degree.
    \item {\tt decs\textunderscore nvss\textunderscore 3m}: Declinations of the $3 ~\rm arcmin$ NVSS matches. The column is formatted as above. Units - degree.
    \item {\tt inds\textunderscore nvss\textunderscore 3m}: Indices of the $3 ~\rm arcmin$ NVSS matches as indexed in the NVSS source catalog. The column is formatted as above.
    \item {\tt fluxes\textunderscore nvss\textunderscore 3m}: Radio flux densities of the $3 ~\rm arcmin$ NVSS matches. The column is formatted as above. Units - mJy.
    \item {\tt flux\textunderscore errors\textunderscore nvss\textunderscore 3m}: Radio flux density errors of the $3 ~\rm arcmin$ NVSS matches. The column is formatted as above. Units - mJy.
    
    \item {\tt num\textunderscore first\textunderscore 30s}: Number of FIRST matches within a $30 ~\rm arcsec$ radius of the galaxy. 
    \item {\tt ras\textunderscore first\textunderscore 30s}: Right Ascensions of the $30 ~\rm arcsec$ FIRST matches. Each object in this column is an array of length five. For the galaxies with less than five $30 ~\rm arcsec$ FIRST matches, dummy values of -999 have been provided. The arrays are sorted in descending order of the $30 ~\rm arcsec$ FIRST matches fluxes. Units - degree.
    \item {\tt decs\textunderscore first\textunderscore 30s}: Declinations of the $30 ~\rm arcsec$ FIRST matches. The column is formatted as above. Units - degree.
    \item {\tt inds\textunderscore first\textunderscore 30s}: Indices of the $30 ~\rm arcsec$ FIRST matches as indexed in the FIRST source catalog. The column is formatted as above. 
    \item {\tt fluxes\textunderscore first\textunderscore 30s}: Radio flux densities of the $30 ~\rm arcsec$ FIRST matches. The column is formatted as above. Units - mJy.

    \item {\tt in\textunderscore erd\textunderscore analysis}: If this value is set to ``True'', galaxy was used for the Eddington ratio distribution analysis of this paper. If this value is set to ``False'', galaxy was not used for the Eddington ratio distribution analysis of this paper.
    \item {\tt is\textunderscore detected\textunderscore agn}: The galaxy hosts a detected radio AGN as defined in Section \ref{ssec:agnid}, if this column is set to ``True''. No detected radio AGN if column is set to ``False''.
    \item {\tt log\textunderscore er}: Logarithm of the Eddington ratio of the AGN. See Section \ref{ssec:ER method} for more information. A dummy value of -999 has been provided for galaxies that aren't classified as detected AGN.
    \item {\tt log\textunderscore sfr\textunderscore sf}: Integrated Star Formation Rate using only the spaxels compatible with recent star formation, provided by the Pipe3D VAC. A dummy value of -999 has been provided for galaxies that do not have valid values of this quantity in the Pipe3D VAC. Units - $\rm \log \left(M_\odot / yr\right)$.
    \item {\tt log\textunderscore ssfr}: Integrated Specific Star Formation Rate, computed using the Pipe3D VAC. A dummy value of -999 has been provided for galaxies that do not have valid values of star formation rate in the Pipe3D VAC. Units - $\rm \log \left(yr^{-1}\right)$.
    \item {\tt log\textunderscore mass}: Stellar mass of the galaxy as provided by the Pipe3D VAC. A dummy value of -999 has been provided for galaxies that do not have valid values of this quantity in the Pipe3D VAC. Units - $\rm \log \left(M_\odot\right)$.
    \item {\tt log\textunderscore stellar\textunderscore sigma\textunderscore 1re}: 1 $R_{e}$ Stellar velocity dispersion, as provided by the DAPall file. A dummy value of -999 has been provided for galaxies that do not have valid values of this quantity in the Pipe3D VAC. Units - $\rm \log \left(km/ s\right)$.
    \item {\tt nsa\textunderscore redshift}: Redshift extracted from the NSA catalog.
    \item {\tt log\textunderscore lr\textunderscore limit}: Upper limit on the radio luminosity of the galaxies without radio matches. This value based on its redshift and NVSS's flux threshold of 2.5 mJy. For galaxies with radio detections, the upper limit is just the measured luminosity. A dummy value of -999 has been provided for galaxies where necessary. Units - $\rm \log \left(ergs/ s\right)$.
    \item {\tt log\textunderscore lr\textunderscore agn\textunderscore thresh}: Threshold radio luminosity that a galaxy would have to reach to be identified as an AGN under our definition (see Section \ref{ssec:agnid}). A dummy value of -999 has been provided for galaxies where necessary. Units - $\rm \log \left(ergs/ s\right)$.

\end{itemize}
\end{document}